\documentclass[submission, Phys]{SciPost}

\usepackage{graphicx}
\usepackage{dcolumn}
\usepackage[active]{srcltx}
\usepackage{color}
\usepackage{comment}
\usepackage{empheq}
\usepackage{tikz}
\usepackage{bigints}
\usepackage{babel}
\usepackage{relsize}
\usepackage{amsmath}
\usepackage{mathtools}
\usepackage{amstext}
\usepackage{amssymb}
\usepackage{amsthm}
\usepackage{amsfonts}
\usepackage{enumerate}
\usepackage{lipsum}
\usepackage[normalem]{ulem}

\usepackage{hyperref}
\hypersetup{
    colorlinks=true,
    citecolor = [rgb]{0.3,0.8,0},
    linkcolor=blue,
    filecolor=magenta,      
    urlcolor=[rgb]{0.4,0.7,0},
    breaklinks=true
}
\usepackage{breakurl}
\newcommand{\HS}[1]{{\color{red}HS: [#1]}}

\theoremstyle{plain}

\theoremstyle{remark}

\newcommand*\widefbox[1]{\fbox{\hspace{2em}#1\hspace{2em}}}
\makeatother

\providecommand{\remarkname}{Remark}
\newtheorem{result}{Result}

\begin{document}

\begin{center}{\Large \textbf{
Nonlocality of Deep Thermalization
}}\end{center}

\begin{center}
Harshank Shrotriya\textsuperscript{1},
Wen Wei Ho\textsuperscript{1,2*}
\end{center}

\begin{center}
{\bf 1} Centre for Quantum Technologies, National University of Singapore, 3 Science Drive 2, Singapore 117543
\\
{\bf 2} Department of Physics, National University of Singapore, Singapore 117542
\\
* wenweiho@nus.edu.sg
\end{center}

\begin{center}
\today
\end{center}


\section*{Abstract}
{\bf
We study the role of global system topology in governing   deep thermalization,
the relaxation of a local subsystem towards a maximally-entropic, uniform distribution of post-measurement states, upon observing the complementary subsystem in a local basis.
Concretely, we focus on a class of (1+1)d systems exhibiting `maximally-chaotic' dynamics, and consider how the rate of the formation of such a universal wavefunction distribution depends on boundary conditions of the system.
We  find that deep thermalization is achieved exponentially quickly in the presence of either periodic or open boundary conditions; however, the rate at which this occurs is twice as fast for the former than for the latter.
These results are attained analytically using the calculus of integration over unitary groups, and supported by extensive numerical simulations. 
Our findings highlight the nonlocal nature of deep thermalization, and clearly illustrates  that the physics underlying this phenomenon goes beyond that of standard quantum thermalization, which only depends on the net build-up of entanglement between a subsystem and its complement. 
}

\vspace{10pt}
\noindent\rule{\textwidth}{1pt}
\tableofcontents\thispagestyle{fancy}
\noindent\rule{\textwidth}{1pt}
\vspace{10pt}

\section{Introduction}

Quantum thermalization refers to the relaxation of a quantum many-body system to a maximally-entropic steady state --- that is, a Gibbs state\footnote{More generally, a generalized Gibbs state, if there are multiple conservation laws.} --- over the course of time. 
This irreversible behavior, which at first sight seems paradoxical to the fact that closed quantum systems undergo unitary and hence reversible time-evolution, pertains more precisely to the behavior of local subsystems: 
over time, there is a build-up of entanglement between a local region and its complement, such that upon ignoring the state of the latter, the former acquires a mixed (and universal) form.
Intuitively, the role of the complementary subsystem, typically assumed to be large compared to the subsystem of interest, is that of a `bath' --- it allows a redistribution of energy, and more generally, any conserved charges. 
Whether or not a system thermalizes \cite{popescu2006entanglement,linden2009quantum,deutsch1991quantum,rigol2008thermalization}, and if so, the timescales it takes, has been the subject of recent intense  studies in the quantum many-body dynamics community, and has given rise to   novel discoveries like ergodicity-breaking mechanisms such as many-body localization \cite{nandkishore2015many,abanin2019colloquium} and quantum many-body scars \cite{serbyn2021quantum,regnault2022quantum}. 

Recently, a new perspective to equilibration under closed quantum many-body dynamics was put forth by Refs.~\cite{proj_ensemble,choi2023preparing}. In this formulation, one is still interested in local properties of a system, however, it is assumed that certain information about the bath is retained --- namely, its classical state as observed by an external agent.
Note that such an assumption is not at all unreasonable in light of current-generation quantum technologies in the form of quantum simulators, which provide microscopically-resolved measurement data through projective measurements of the  global system.
Thus, one can construct a hybrid quantum-classical description of a local subsystem in terms of an ensemble of  pure (quantum)  states, each of which is conditioned upon a particular (classical) measurement outcome of the bath. 
This constitutes the so-called {\it projected ensemble}. Recent works have demonstrated, empirically \cite{proj_ensemble}, experimentally \cite{choi2023preparing} and then later rigorously in a number of exactly-solvable models \cite{ho2022exact,ippoliti2023dynamical}, that in certain cases, the distribution of the projected ensemble tends towards a universal form in which the pure states are uniformly (and hence maximally-entropically) distributed over the Hilbert space they live on. This phenomenon was dubbed {\it deep thermalization}, as it is a form of equilibration more stringent than regular thermalization
--- it occurs at the level of individual wavefunctions underlying the local subsystem, and not at the level of local observables. 
This difference can be elucidated quantitatively by asking about the times it takes to achieve regular (deep) thermalization, i.e., the time taken for a local subsystem to reach a maximally-entropic mixed state (wavefunction distribution).
%
In the exactly-solvable models alluded to above, it was found that regular and deep thermalization times precisely match \cite{ho2022exact, ippoliti2023dynamical}. However, in Ref.~\cite{ippoliti2023dynamical} it was argued that this was the result of the fine-tuned nature of these models underpinning their exact-solvability; more generally, it is expected that there will be a difference between these times, leveraging results of dynamical purification in monitored quantum circuits for (1+1)d systems. Subsequent work \cite{ippoliti2022solvable} demonstrated a model where it can be analytically shown that these times are indeed gapped. 

In the present work, we investigate the role of topology of the global system in governing  regular and deep thermalization times.
Precisely, we focus on a small, local region located deep in the bulk of a thermodynamically large quantum many-body system, and investigate the rate of how it equilibrates depending on the connectivity of the boundaries of the system.
For unitary dynamics generated by Hamiltonians with a sense of geometric locality,  bounds on information propagation bounds (\`a la Lieb and Robinson \cite{lieb2004finite}) restrict the build-up of entanglement of a spatially local region to be with a region within a finite light-cone surrounding it. Thus, dynamics of local observables and hence  regular thermalization times are effectively independent of the  global topology of the rest of the system.  
In contrast, deep thermalization entails a measurement of the bath, a dynamical process which evades information propagation bounds: thus, spatially distinct parts of the system, even those outside the light-cone, may drastically affect each other. 
Studying different connectivities of a system therefore offers an avenue to  highlight the potentially distinct physics that underpin these two different notions of local equilibration. 

To make our investigations concrete,  we focus on the emergence of deep thermalization in a model describing a periodically-driven 1d array of spin-1/2s, the kicked Ising model (KIM). 
The KIM constitutes one of the models where deep thermalization has been rigorously proven to occur \cite{ho2022exact}, wherein the region of interest constitutes a small subsystem of finite size $N_A$ located at one end of a thermodynamically large chain.  Our present work differs from Ref.~\cite{ho2022exact}  by considering a region of interest located in the {\it bulk} of a large chain, with the left and right ends of the system either connected with periodic boundary conditions (PBC) or with open boundary conditions (OBC). 
We find, analytically using the calculus of integrals over the unitary group as well as supporting numerics, that in both cases regular thermalization is achieved in this model {exactly} at time $t = \lceil{N_A/2\rceil}$, while deep thermalization requires taking the additional limit of large times $t \to \infty$ (this is in contrast to the set-up considered in Ref.~\cite{ho2022exact}, where regular and deep thermalization {\it both} occur exactly at $t = N_A$). 
Intriguingly, the rates of deep thermalization   depend crucially on the topology of the system: while in both cases the convergence to a locally maximally-entropic wavefunction distribution is exponentially fast, for a system with PBC, deep thermalization is achieved at a rate {\it twice} that in a system with OBC. In other words, the deep thermalization time in a system with OBC is twice as long as in a system with PBC. 
%
%

Our findings 
illustrate unambiguously deep thermalization's nonlocal nature,
and show that it 
is underpinned by novel physics beyond that of standard quantum thermalization.
Indeed, our results demonstrate that the projected ensemble is sensitive not  only to the  net build-up of   entanglement between a subsystem and its complement, but also to the internal structure of how this entanglement is organized (probed by local measurements). Thus, the projected ensemble and the emergence of deep thermalization possibly serves as a probe of other interesting features of  dynamics that are sensitive to the fine-grained details of entanglement generation, such as quantum information scrambling. 


The rest of the paper is organized as follows. In Sec.~\ref{sec:setup} we quickly recap the framework of projected ensembles and  the physical phenomenon of deep thermalization. 
We explain how deep thermalization can be quantitatively  studied using the notion of  quantum state-designs from quantum information theory, allowing us to define deep thermalization   times. We then introduce the kicked Ising model (KIM), the system studied in this work to understand the role of different topologies in determining   deep thermalization times. 
Sec.~\ref{sec:analytics} presents our analysis of the KIM, utilizing a tensor network representation of dynamics as well as integral calculus over Haar random unitary matrices (i.e., Weingarten calculus). We show how the rates of regular and deep thermalization depend sensitively on the connectivity of the system considered. 
Sec.~\ref{sec:numerics} supports our analytical findings with numerical simulations, where we find very good agreement with our theoretical predictions.
We   conclude with a summary and discussion of our findings in Sec.~\ref{sec:discussion}, and highlight some open directions for future work.

\section{Setup}
\label{sec:setup}
We first  review the projected ensemble, referring  the reader to the original works,  Ref.~\cite{proj_ensemble, choi2023preparing, ho2022exact} for a more detailed exposition and motivation of its construction.
We then describe the  nonequilibrium phenomenon that emerges within this setting known as deep thermalization, before presenting the kicked Ising model (KIM), the system of interest considered in this paper.

\subsection{Preliminaries}
\subsubsection{The projected ensemble}

The projected ensemble is constructed as follows. Consider a pure quantum state $|\Psi\rangle_{AB}$ defined on a bipartite many-body system $A \cup B$. This could, for example, be an energy eigenstate of the   Hamiltonian or an entangled state arising in quench dynamics.
For simplicity, we may assume that we are  considering a system of $N = N_A + N_B$ qubits, though this assumption can be lifted to allow for systems of qudits or fermions of arbitrary (but finite) local dimension.

\begin{figure}[t]
  \includegraphics[width=\columnwidth]{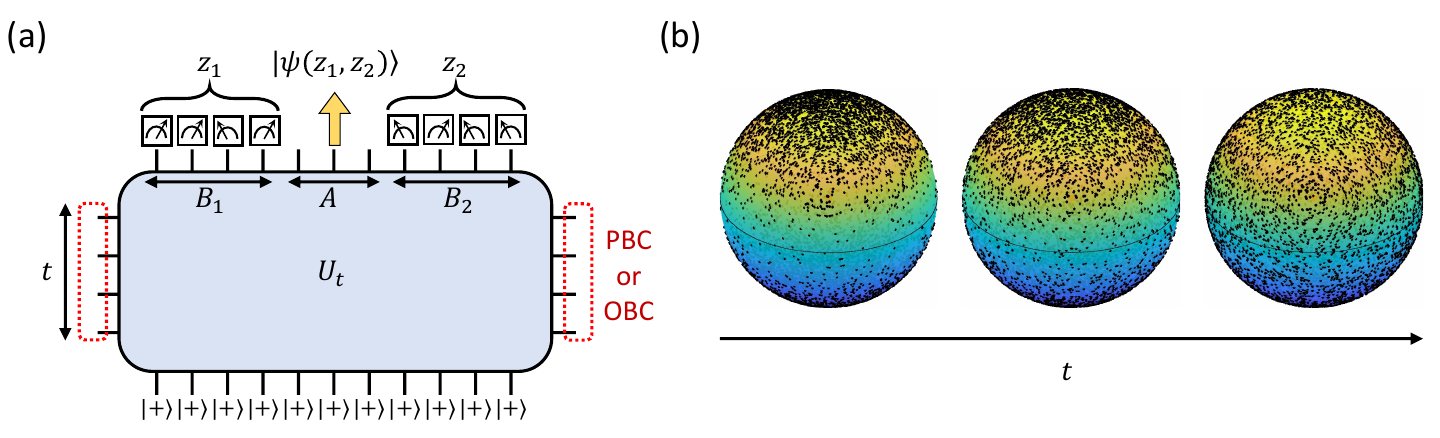}
  \caption{(a) Construction of the projected ensemble. The ensemble consists of pure projected states $|\psi(z)\rangle$ of a local subsystem $A$, which are post-measurement (conditional) states upon measuring the complementary region (`bath') $B$ in a complete basis (e.g., the computational-$z$ basis) and obtaining measurement outcome $z$. In this work we are concerned with the projected ensemble arising from dynamics under unitary time evolution $U_t$, with $  A$ being a small region located in the bulk, flanked by `baths' $B_1$ and $B_2$ on the left and right, and connected with either periodic boundary conditions (PBC) or open boundary conditions (OBC). Thus, the measurement outcome $z$ is $z = (z_1, z_2)$. 
  (b) Cartoon of deep thermalization and difference with regular thermalization. For a local region $A$ being a single qubit, we can plot the distribution of projected states of the projected ensemble as points on the Bloch sphere. Deep thermalization refers to the attainment of a uniform distribution over Hilbert space over time. 
  Note that regular thermalization does not distinguish between the three distributions shown, even though the left-most distribution is clearly not uniform; in all cases, the barycenter (mean) of all the projected states is a point in the middle of the Bloch sphere, which corresponds to regular quantum thermalization to infinite temperature.
  }
  \label{fig_1}
\end{figure}

Suppose now we perform a projective measurement on $B$ in a complete basis $\{ |z\rangle \}$, with measurement outcomes labeled by $z$. Then, the state on $B$ will be updated to $|z\rangle$, while the state on $A$ will be in a definite {\it pure} state\footnote{It is pure because we are assuming that the information obtained on the bath is {\it complete}, that is, $\{ |z\rangle\langle z| \}$ are orthogonal rank-1 projectors that span the Hilbert space of $B$.} given by the Born update: \begin{align}
    |\psi(z)\rangle_A := (\mathbb{I}_A \otimes \langle z|_B) |\Psi\rangle_{AB}/\sqrt{p(z)},
    \label{eqn:proj_state}
\end{align}
where $p(z) =  \langle \Psi|_{AB} ( \mathbb{I}_A \otimes |z\rangle_B\langle z| ) |\Psi\rangle_{AB}$ is the probability that outcome state $z$ is observed, see Fig.~\ref{fig_1}(a). 
Motivated by experimental considerations (see  \cite{proj_ensemble, choi2023preparing}), we are typically interested in the case where $z \in \{0, 1\}^{N_B}$ is a bit-string that represents a local `classical' configuration of the system, where each bit provides information, for example, of the position of a particle in an optical lattice, or the internal spin state of  an atom or ion in a given trap. In quantum computing parlance, measuring in the basis $\{ |z\rangle\}$ thus corresponds to a `computational basis' measurement, the natural way of extracting information in a many-body system.
Considering   the conditional projected states $|\psi(z)\rangle_A$ and their respective probabilities $p(z)$ over all possible states of the bath $z$ then yields an ensemble of pure states called the projected ensemble,
\begin{equation}
         \mathcal{E} = \{(p(z), |\psi(z)\rangle_{A}): z \in \{0,1\}^{N_B}\},
        \label{eqn:PE}
\end{equation}
which can be understood as a {\it distribution} of wavefunctions over the Hilbert space\footnote{More precisely, the projected ensemble is a distribution over the complex  projective space.} $\mathcal{H}_A$ of $A$.

Information captured by the projected ensemble includes, for example, {\it conditional} 1-point expectation values of local observables, 
$
    \langle O_A \rangle_z \equiv \text{Tr}_{A} \left( O_A |\psi(z)\rangle_A \langle  \psi(z)| \right) = \text{Tr}_A \left( O_A \rho_A(z) \right),
$
i.e., the   expectation value of a quantum observable $O_A$ on local region $A$ conditioned upon the bath $B$  being in state $z$. 
Averaged over all states of the ensemble, the mean conditional expectational value is 
$\mathbb{E}_z [ \langle O_A \rangle_z ] = \sum_z p(z) \text{Tr}_{A} \left( O_A |\psi(z)\rangle_A \langle  \psi(z)| \right) = \text{Tr}_A \left( O_A \rho_A \right)$, where $\rho_A = \text{Tr}_B( |\Psi\rangle_{AB} \langle \Psi|_{AB} )$ is the reduced density matrix (RDM), which we recognize to be the standard expectation value of the observable $O_A$  {\it agnostic} of the state of the bath. This thus shows that the projected ensemble $\mathcal{E}$ can recover any information that the RDM $\rho_A$ contains. Examples of information captured by $\mathcal{E}$ but not $\rho_A$ include conditional $k$-point functions $\mathbb{E}_z [ \langle O_{1,A} \rangle_z \langle O_{2,A} \rangle_z \cdots \langle O_{k,A} \rangle_z  ]$ for any $k \geq 2$ (a subset of which has in fact been measured by \cite{choi2023preparing}), 
which demonstrates that the projected ensemble is a novel quantum-classical description of a local subregion of a quantum many-body system which generalizes the reduced density matrix formalism  as it  retains, rather than discards, certain correlations with its environment.

\subsubsection{Deep thermalization}
Consider now a global many-body state $|\Psi\rangle_{AB}$ which arises from quench dynamics under a quantum chaotic Hamiltonian, beginning from a nonequilibrium initial configuration. 
Conventional wisdom (specifically, the second law of thermodynamics) informs us that   the reduced density matrix $\rho_A(t)$   of a small system $(N_A \ll N_B)$ maximizes its entropy over time, subject to conservation laws, which implies it relaxes to a universal  mixed Gibbs state: $\rho_A(t) \to \approx e^{-\beta H_A}/Z$ ($\beta$: inverse temperature, $Z$: normalization corresponding to the partition function)  --- i.e., the system quantum thermalizes.
It is thus natural to inquire if the  projected ensemble $\mathcal{E}$ itself    also relaxes  towards a universal distribution.

In the case of chaotic dynamics in the absence of any conservation laws where the reduced density matrix is expected to tend towards a  featureless maximally-mixed state $\rho_A(t) \to \mathbb{I}_A/2^{N_A}$, it seems reasonable to imagine that the projected ensemble similarly tends towards a  featureless, maximally-entropic distribution in Hilbert space, a phenomenon that can be dubbed {\it deep thermalization}. This was first conjectured by Refs.~\cite{proj_ensemble,choi2023preparing}.
Mathematically, this amounts to asking if the projected ensemble $\mathcal{E}(t)$, generated by a time-evolving state $|\Psi(t)\rangle_{AB}$, tends in time to a limiting form given by an ensemble of uniformly distributed states
\begin{align}
    \mathcal{E}_\text{Haar} = \{ dU_A, |U_A\rangle \},
\end{align}
where $dU_A$ is the Haar measure on the space of unitaries on $A$ and $|U_A\rangle := U_A|\phi_0\rangle$ for any arbitrary but fixed generator state $|\phi_0\rangle$.
For this reason we shall refer to the corresponding collection of states $\mathcal{E}_\text{Haar}$ as the Haar ensemble. 
Indeed, Refs.~\cite{proj_ensemble,choi2023preparing} provided strong empirical   and experimental  evidence that such universality arises; this was later substantiated by rigorous results demonstrating  its emergence  in a family of exactly-solvable models \cite{ho2022exact,ippoliti2023dynamical}, which includes the periodically-kicked Ising model, and quantum circuits made up of special local quantum gates which possess a so-called dual-unitary property \cite{akila2016particle,bertini2018exact,gopalakrishnan2019unitary,claeys2020maximum}. Such discoveries   hint at a possible generalized guiding principle beyond the second law of thermodynamics governing the ultimate fate of quantum many-body systems over time, wherein the notion of an entropy that is maximized should be appropriately generalized (for example, from the von Neumann entropy of the reduced density matrix to possibly the Shannon entropy of the projected ensemble). 
For the special case of  free-fermionic (and hence non-chaotic) systems which possess infinitely many integrals of motion, recent work \cite{lucas2023generalized} has conjectured and shown numerically that the 
limiting wavefunction distribution in dynamics is consistent with the projected ensemble obtained from a representative Gaussian state whose conserved charges match those of the initial state. As this is a generalization of equilibration of the reduced density matrix towards the (non-thermal) generalized Gibbs ensemble (GGE), the authors dubbed this `generalized deep thermalization'.

Returning to chaotic quantum many-body dynamics without conservation laws, the degree of deep thermalization can be systematically quantified by  probing whether the projected ensemble $\mathcal{E}$ matches that of the uniform, Haar ensemble $\mathcal{E}_\text{Haar}$ at moment $k$ of the respective distributions.
Precisely, for the $k$-th moment we may form the following density operators on the $k$-replicated Hilbert space $\mathcal{H}_A^{\otimes k}$ of $A$ 
\begin{align}
     \rho^{(k)} &:= \sum_{z=1}^{N_B} p(z) ( |\psi(z)\rangle_A \langle \psi(z)|_A)^{\otimes k}, \label{eqn:rho_k} \\
    \rho^{(k)}_\text{Haar} &:= \int dU_A (|U_A\rangle \langle U_A|)^{\otimes k};
    \label{eqn:rho_Haar_k}
\end{align}
these operators neatly package any conditional $k$-point function $O_1 \otimes O_2 \otimes \cdots \otimes O_k$  within the  corresponding ensemble:
\begin{align}
     \mathbb{E}_{\mathcal{E}} [ \langle O_{1} \rangle_z \langle O_{2} \rangle_z \cdots \langle O_{k} \rangle_z  ] &= \text{Tr}\left( \rho^{(k)} O_1 \otimes O_2 \otimes \dots \otimes O_k \right), \\
     \mathbb{E}_{\mathcal{E}_\text{Haar}} [ \langle O_{1} \rangle_{U_A} \langle O_{2} \rangle_{U_A} \cdots \langle O_{k} \rangle_{U_A}  ] &= \text{Tr}\left( \rho^{(k)}_\text{Haar} O_1 \otimes O_2 \otimes \dots \otimes O_k \right).
\end{align}
If all such expectation values agree between the ensembles up to some small precision $\varepsilon > 0$ (with operators appropriately normalized), or equivalently if the density operators' trace distance obey
\begin{equation}
    \Delta^{(k)} \coloneqq \frac{1}{2} \Vert \rho^{(k)} - \rho_{\text{Haar}}^{(k)} \Vert_1 \leq \varepsilon,
\end{equation}
then we say that the system has (approximately) deep thermalized at level $k$. 
Note that approximate deep thermalization at level $k$   implies approximate deep thermalization at level $k-1$ but not vice versa.
In particular, a system thermalizing to infinite temperature at the level of the reduced density matrix ($\Delta^{(k=1)} = 0$) does not imply the projected states are uniformly distributed in Hilbert space (captured by $k\geq2$ moments), demonstrating that deep thermalization is more stringent than regular thermalization, see Fig.~\ref{fig_1}b.
The equivalence of $\mathcal{E}$ to $\mathcal{E}_\text{Haar}$ at the $k$-th moment is commonly referred to in quantum information theoretical language as forming a quantum state $k$-design \cite{divincenzo2002quantum,ambainis2007quantum,gross2007evenly}.


Using this notion of convergence, we can, in the case of a projected ensemble formed in quench dynamics beginning from a nonequilibrium initial state,   define a sequence of {\it design times} $t_k$, as the minimum time it takes for the projected ensemble to fall below some fixed threshold, i.e., 
\begin{align}
    t_k := t \text{~~s.t.~~} \Delta^{(k)}(t) \leq \varepsilon.
    \label{eqn:deep_thermalization_times}
\end{align} Note that these times are monotonic, $t_{k+1} \geq t_k$.
The first design time $t_1$ --- the time for the RDM to approximate the maximally mixed state, is nothing more than the regular thermalization time, while higher design times $t_{k \geq 2}$ correspond to progressive levels of deep thermalization; we thus refer to the latter collectively as the {\it deep thermalization} times.
Differences in $t_k$ for different $k$ can signal distinct physics at play, which prior works Refs.~\cite{ippoliti2023dynamical, ippoliti2022solvable} have studied.
In this work, we shall be concerned with the differences between regular and deep thermalization times that arise due to   different topologies of a system. 


\subsection{Model}
The system we consider consists of a periodically-kicked 1D chain of $N$ spin-$1/2$ particles (i.e., qubits) called the kicked Ising model (KIM). 
Concretely, dynamics is generated by repeated applications of the Floquet unitary operator
\begin{equation} \label{floquet_unitary}
    U_F = U_h e^{-i H_{\textrm{Ising}} \tau},
\end{equation}
 which entails periodic alternation between an entanglement-generating Ising Hamiltonian $H_\text{Ising}$ applied for time $\tau = 1$ and a transverse kick $U_h = \exp(-i h \sum_{i=1}^{N}\sigma_i^{y})$ with transverse field strength $h$.
We shall be interested in chains defined with two different connectivities --- periodic boundary conditions (PBC) and open boundary conditions (OBC), so that the Ising Hamiltonian reads:
\begin{align}
    H_{\text{Ising}} = \begin{cases} J \sum_{i=1}^{N}\sigma_i^z \sigma_{(i ~ \text{mod} ~ N) + 1}^z + g\sum_{i=1}^N \sigma_i^z  , & \text{(PBC)}\\
    \\
    J \sum_{i=1}^{N-1}\sigma_i^z \sigma_{i+1}^z + g\sum_{i=1}^N \sigma_i^z +b_1\sigma_1^z + b_N\sigma_N^z, & \text{(OBC)}
  \end{cases}
\end{align} 
where $J$ is the strength of Ising interactions, $g$ is the strength of a longitudinal field, and $b_1, b_N$ are boundary fields. We fix both to have value $\pi/4$; they are introduced solely to simplify technical simplifications and are irrelevant for the ensuing physics.

For generic values of parameters $(J,g,h)$ --- specifically, avoiding the free fermionic and Clifford points of the model, the KIM is believed to generate quantum chaotic dynamics. Indeed, in the special case of $(|J|,|h|) = (\pi/4, \pi/4)$, it has been shown that its spectral statistics are exactly captured by predictions from random matrix theory 
\cite{bertini2018exact}. For this reason the KIM with these parameters is viewed as  `maximally chaotic'.
In the context of the projected ensemble, for similar parameters 
$(|J|,|h|) = (\pi/4, \pi/4)$ and any $g$ excluding the points $\mathbb{Z}\pi/8$, and beginning from the $x$-polarized initial state $|+\rangle^{\otimes N}$,
both regular and deep thermalization (at any $k$) were proven to {\it exactly} occur for a subsystem $A$ defined to be a contiguous region of $N_A$ spins located on one end of a thermodynamically large system with OBC, in time $t = N_A$ \cite{ho2022exact}. In other words, all design times collapse: $t_k = N_A$. The basic picture uncovered was that the projected states can be viewed as arising from   stochastic `quantum computations' through the bulk of the system with measurements indexing the   computation, upon performing a space-time duality transformation of the quantum circuit underlying the system. This is similar to the computation that arises through the entanglement present between different spatial regions of resource states in measurement-based quantum computation (see \cite{stephen2022universal}). Further, these computations were shown to be universal, in the sense that they map to every point in Hilbert space, hence leading to an emergent uniform distribution within the projected ensemble. More details can be found in Ref.~\cite{ho2022exact}. 



In our present work, we again consider the many-body state 
\begin{equation}
    |\Psi(t)\rangle = U_F^t |+\rangle^{\otimes N}
\end{equation}
arising from $t \in \mathbb{N}$ applications of $U_F$ on the initial state $|+\rangle^{\otimes N}$ with similar system parameters as Refs.~\cite{ho2022exact}, but now construct the projected ensemble $\mathcal{E}$ on $A$ consisting of $N_A$ contiguous spins located in the {\it middle} of the system, namely centered around site $\lfloor N/2 \rfloor$, for both PBC and OBC. The complementary subsystem $B$ is thus such that $B=B_1 \cup B_2$, where $B_1 (B_2)$ are regions to the left (right) of region $A$, see Fig.~\ref{fig_1}(a).
The projected ensemble $\mathcal{E}$, Eq.~\eqref{eqn:PE}, is therefore composed of projected states still given by Eq.~\eqref{eqn:proj_state}, but now with the classical configuration of the bath labeled by $z = (z_1, z_2)$, where $z_{1} = \{0,1\}^{N_{B_1}}$ and $z_{2} = \{0,1\}^{N_{B_2}}$, which represent the bit-string outcomes of a joint projective measurement on regions $B_1$ and $B_2$ respectively. 

In this scenario, prior work on entanglement growth   has demonstrated that the von Neumann entropy $S(t)$ grows linearly at early  times with a slope $v = 2$ (measured in log base 2), and saturates at time $t_1 = \lceil N_A/2\rceil$ \cite{bertini2019entanglement} to $S(t) = N_A$. Thus $\rho_A$ becomes exactly equal to the maximally mixed state $\mathbb{I}_A/2^{N_A}$ at that time. This constitutes regular quantum thermalization to infinite temperature. However, deep thermalization has not yet been  established for this setup, which we proceed to do next. 
As regular quantum thermalization is a necessary condition for deep thermalization to occur, we henceforth concentrate on times $t \geq \lceil N_A/2 \rceil$.

\section{Analytical solution}
\label{sec:analytics}
\subsection{Projected ensemble in the thermodynamic limit}

We start our analysis by recognizing that the dynamics under the Floquet unitary $U_F$ admits a quantum circuit representation which can  be cast as a tensor network diagram. This representation was already introduced and underlies the proof of the result of Ref.~\cite{ho2022exact} (see also \cite{stephen2022universal} where it was used), so below we simply quickly summarize the relevant key basic tensors and rules underlying the representation\footnote{These rules are reminiscent of so-called ZX calculus, a graphical language for reasoning about quantum computation, see for example Ref.~\cite{coecke2008interacting,coecke2011interacting}.}.
By using such manipulations on the basic diagrams, we will derive a simplified form of the projected ensembles for both the PBC and OBC cases in the thermodynamic limit, Eqs.~\eqref{eqn:EPBC}, \eqref{eqn:EOBC}, upon invoking a space-time duality transformation. 

We first introduce the following elementary tensors
\begin{equation}
     \raisebox{-.4\height}{\includegraphics[height=0.7cm]{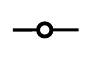}} = \frac{1}{\sqrt{2}} \begin{pmatrix}
  1 & 1\\ 1 & -1 \end{pmatrix}, \qquad\qquad \raisebox{-.4\height}{\includegraphics[height=1.25cm]{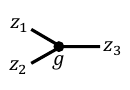}} = \delta_{z_1 z_2 z_3}e^{-ig(1-2z_1)},
\end{equation}
where each leg is labeled by an index $z_i \in \{0,1\}$, 
which are tensor network representations of the Hadamard gate $H$ (whose action is $H|0\rangle = |+\rangle$ and $H|1\rangle = |-\rangle$), and a 3-legged tensor specified by parameter $g$, which consists of a 3-legged Kronecker delta   ($\delta_{z_1 z_2 z_3} = 1$ if $z_1 = z_2=z_3$ and $0$ otherwise) together with a  phase factor ($e^{- i g} (e^{+i g})$ if the inputs are $0(1)$).
Using these   we can, ignoring irrelevant global phases, cast evolution under the Ising interaction between two qubits  and  a local transverse kick (for the parameters  chosen) respectively as,
\begin{equation}
    e^{-i\frac{\pi}{4}\sigma^{z}\otimes \sigma^{z}} = \raisebox{-.4\height}{\includegraphics[height=1.65cm]{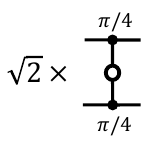}}, \qquad\qquad e^{-i\frac{\pi}{4}\sigma^{y}} = \raisebox{-.4\height}{\includegraphics[height=1.00cm]{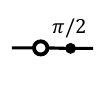}}.
\end{equation}
Above, we read the action of the tensors from right (i.e., input) to left (i.e., output). 
Some basic rules of the tensors include contraction of two Kronecker deltas and contraction with a $|+\rangle$ state,
\begin{equation}
    \raisebox{-.4\height}{\includegraphics[height=1.25cm]{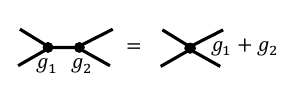}}, \qquad\qquad \raisebox{-.45\height}{\includegraphics[height=1.25cm]{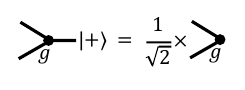}}.
\end{equation}
Additionally, a measurement in the computational basis at site $i$ in region $B_1$ yields a contraction with state $|z_{B_1,i}\rangle$ 
\begin{align} 
    \raisebox{-.5\height}{\includegraphics[height=1.0cm]{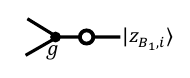}} = \begin{cases} \raisebox{-.45\height}{\includegraphics[height=1.cm]{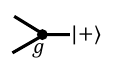}}  & \text{if $z_{B_1,i} = 0$},\\
    \raisebox{-.5\height}{\includegraphics[height=1.1cm]{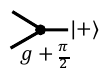}} & \text{if $z_{B_1,i} = 1$},
  \end{cases}
\end{align}
and similarly for sites in $B_2$.

\begin{figure}[hbt!]
  \includegraphics[width=\textwidth]{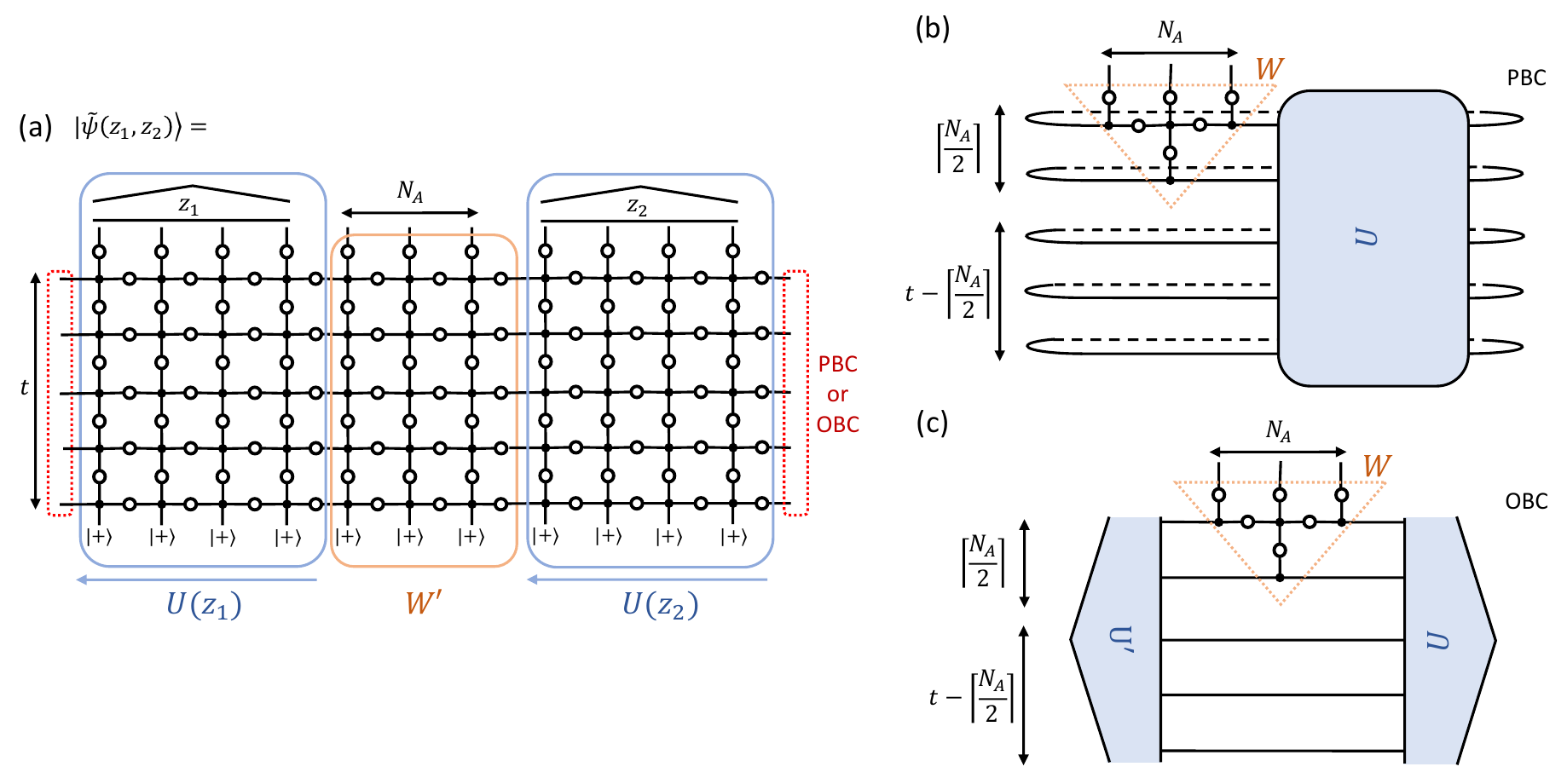}
  \caption{(a) Tensor network representation of the unnormalized projected state $|\tilde{\psi}(z_1,z_2)\rangle$ on subsystem $A$, Eq.~\eqref{unnorm_state}. The blue boxes define unitaries $U({z_1}), U({z_2})$ up to a proportionality factor; while $W'$ is defined by the tensors in the orange box as a linear map from two temporal $t$ slices to the spatial region $A$.   
  (b,c) Simplified tensor network representations of the state upon taking the thermodynamic limit $N_{B_1} \rightarrow \infty$ and $N_{B_2} \rightarrow \infty$. 
  (b) PBC case. Resulting tensor network representation of an  unnormalized projected state, where the map $W'$ can be simplified to $W$ (defined by the orange triangle), and where the contributions of $U(z_1), U(z_2)$ can be combined into a single unitary $U$, which is traced over together with the inserton of $W$. 
  In the projected ensemble the corresponding normalized state is sampled with probability proportional to the unnormalized state's squared norm, where unitary $U$ is drawn randomly from the Haar measure $dU$ on $t$ qubits. 
  (c) OBC case. 
  Resulting tensor network representation an unnormalized projected state, where $W$ is the same as in the PBC case.
  However, now instead of a single random unitary which is traced over, the unnormalized projected state involves contraction by two states $|U\rangle$ and $|U'\rangle$.
  The probability that the corresponding normalized state occurs in the projected ensemble is proportional to the unnormalized state's squared norm, where $|U\rangle = U|+\rangle^{\otimes t}, |U'\rangle = U' |0\rangle^{\otimes t}$ are random states drawn by sampling $U,U'$ from the Haar measure $dU, dU'$ on $t$ qubits, independently.
  }
  \label{fig_2}
\end{figure}

Therefore up to a common multiplicative factor, an unnormalized projected state $|\tilde{\psi}(z_1,z_2)\rangle = \langle z_1, z_2| U_F^t |+\rangle^{\otimes N}$ on subsystem $A$ can  be diagrammatically depicted as shown in Fig.~\ref{fig_2}a, where the horizontal direction represents space (precisely, of $N$ qubits) and the vertical direction represents time (precisely, of $t$ applications of $U_F$). The left and right ends of the diagram are terminated either by connecting the dangling legs together (PBC), or contracted on the left and right with the states $\langle +|^{\otimes t}$ and $|0\rangle^{\otimes t}$ respectively (OBC). 
On the top of the diagram, there is a partial contraction of the tensor network in the spatial region $B_1$ with the product state $\langle z_1| = \otimes_{\text{sites } i \in B_1} \langle z_i|$, and similarly in the region $B_2$ with the  product state $\langle z_2| = \otimes_{\text{sites } i \in B_2} \langle z_i|$, according to the measurement outcomes $(z_1, z_2)$ on the bath $B = B_1 \cup B_2$.

One sees from Fig.~\ref{fig_2}(a) that away from the boundaries, the bulk of the tensor network diagram looks identical when viewed from bottom to top or right to left.
As the tensor network describes unitary evolution vertically, this means it also describes unitary evolution horizontally --- this special property is known as {\it dual-unitarity}\footnote{More specifically, the KIM is actually {\it self-dual}, since not only is it unitary in both the space and time directions, the unitary evolution is the same.}, and   holds for the model dialled to the special parameters values we are considering. 
Now, the partial contraction of the bulk tensor network on the bottom boundary with the input state $|+\rangle^{\otimes N}$   and  the top boundaries with the computational basis states $\langle z_1|_{B_1} \otimes \langle z_2|_{B_2}$ respectively allows one to define two maps $U({z_1}), U({z_2})$ acting on the space of $t$ qubits (see blue boxes in Fig.~\ref{fig_2}a for their precise definitions, reading their action from right to left), and it turns out that these are in fact {\it unitaries}. That is, $U({z_1})^\dagger U({z_1}) = U({z_2})^\dagger U({z_2}) = \mathbb{I}$. We refer the reader to Ref.~\cite{ho2022exact} which contains more details of the construction of such objects.
Similarly, we may also define a multi-linear map $W'$ (orange box in Fig.~\ref{fig_2}(a)), which maps the space of two copies of $t$ qubits (`temporal slices') to the Hilbert space $\mathcal{H}_A$ of $N_A$ qubits (`spatial region'); this tensor has indices $W'^\sigma_{\tau \tau'}$ where the spatial index $\sigma \in \{0,1\}^{N_A}$ and temporal indices $\tau,\tau' \in \{0,1\}^{t}$.


Thus, one has a useful dual interpretation of a projected state as arising from `quantum evolutions' of $t$ temporal qubits effected by $U({z_1})$ and $U({z_2})$, with the subsequent operation $W'$ then mapping such information to the spatial region of $N_A$ qubits.  Precisely, we may express a projected state $|\Tilde{\psi}(z_1,z_2)\rangle$ as
\begin{align} \label{unnorm_state}
    |\Tilde{\psi}(z_1,z_2)\rangle \propto \begin{cases} 
    \sum_{\sigma \in \{0,1\}^{N_A} } \textrm{Tr}\left( U({z_1}) W'^{\sigma} U({z_2}) \right) |\sigma \rangle, & \text{(PBC)},\\
    \\
    \sum_{\sigma \in \{0,1\}^{N_A} } \langle +|^{\otimes t} U({z_1}) W'^{\sigma} U({z_2}) |0\rangle^{\otimes t} |\sigma \rangle, & \text{(OBC)},
  \end{cases}
\end{align}
where the trace is over the temporal space, whose indices we have suppressed. 
Note the probabilities $p(z_1,z_2)$ in the projected ensemble are proportional to the overlap  $\langle \tilde{\psi}(z_1,z_2)|\tilde{\psi}(z_1,z_2)\rangle$ of the unnormalized projected states.


While the above diagrammatic manipulations are rather detailed,     simplifications arise   upon taking the thermodynamic limit (TDL) $N_{B_1}, N_{B_2} \to \infty$. 
It was shown (Theorem 2) in Ref.~\cite{ho2022exact} that the unitary ensembles $\{U({z_1})\}$ and $\{U({z_2})\}$, indexed by measurement outcomes $z_1$ and $z_2$ respectively and with each element considered equally likely, are statistically indistinguishable from Haar random unitaries over $t$ qubits. 
Mathematically, this is captured by the following statements:
\begin{align} \label{tdl_limit}
   & \lim_{N_{B_1} \rightarrow \infty} \sum_{z_1 \in \{0,1\}^{N_{B_1}}  } \frac{1}{2^{N_{B_1}}} U({z_1})^{\otimes k} \otimes U({z_1})^{*\otimes k} =  \int_{U \sim \textrm{Haar}(2^t)} dU U^{\otimes k} \otimes U^{* \otimes k}, \nonumber \\
   & \lim_{N_{B_2} \rightarrow \infty} \sum_{z_2 \in \{0,1\}^{N_{B_2}}  } \frac{1}{2^{N_{B_2}}} U({z_2})^{\otimes k} \otimes U({z_2})^{*\otimes k} = \int_{U \sim \textrm{Haar}(2^t)} dU U^{\otimes k} \otimes U^{* \otimes k}.
\end{align}

Therefore, we can replace every instance of $U(z_1) (U(z_2))$ with a Haar random unitary $U (U')$ on $t$ qubits, so that the projected ensemble in the thermodynamic limit acquires the following limiting form:
\begin{equation}
    \mathcal{E} = \left\{dU  dU' p(U,U') , |\psi(U,U')\rangle = \frac{|\Tilde{\psi}(U, U')\rangle}{\Vert |\Tilde{\psi}(U, U')\rangle \Vert} \right\},
\end{equation}
where $p(U,U') \propto \Vert |\Tilde{\psi}(U, U')\rangle \Vert^2$, normalized appropriately so that $\int\int dU dU' p(U,U') = 1$ with $dU, dU'$ the Haar measure on the space of unitaries acting on $t$ qubits, and with the unnormalized projected state given by
\begin{align} \label{unnorm_1}
    |\Tilde{\psi}(U, U')\rangle \propto \begin{cases} \sum_{\sigma \in \{0,1\}^{N_A} } \textrm{Tr}\left( U W'^{\sigma} U' \right) |\sigma \rangle & \text{(PBC)}, \\\\
    \sum_{\sigma \in \{0,1\}^{N_A} } \langle +|^{\otimes t} U W'^{\sigma} U' |0\rangle^{\otimes t} |\sigma \rangle & \text{ (OBC)}.
  \end{cases}
\end{align}

Finally we perform one more round of simplifications, utilizing the left and right invariance of the Haar measure.
First, the tensor $W'$ can be reduced to the triangular tensor $W$ (orange triangle of Fig.~\ref{fig_2}(b,c) for its pictorial definition; though there is a slight difference in its construction for even versus odd $N_A$s, see Appendix \ref{sec:W_tensor} for details) by pulling out two unitaries on the left and right which can be absorbed into $U$ and $U'$ respectively. $W$ is defined to be a map from two temporal spaces of ${\lceil N_A/2 \rceil}$ qubits to the Hilbert space $\mathcal{H}_A$ of $N_A$ qubits, with matrix elements $W^\sigma_{\tau,\tau'}$, where $\sigma \in \{0,1\}^{N_A}$ and $\tau,\tau' \in \{0,1\}^{\lceil N_A/2 \rceil}$ (recall we are working at times $t \geq \lceil N_A/2 \rceil$). 
Straightforwardly, but importantly, one can also check from the diagrammatic rules that this tensor is proportional to an isometry: $\sum_{\tau \tau'} (W^\sigma_{\tau \tau'} W^{* \sigma'}_{\tau \tau'}) \propto \delta_{\sigma \sigma'}$, see Appendix 
 \ref{sec:W_tensor_contraction}. Second, we  note that in the PBC case, the unitaries $U$ and $U'$ can in fact be combined into a single unitary and one of the measures integrated over. 

Therefore, we  end up with the following final form of the projected ensemble in the TDL:
\begin{empheq}[box=\widefbox]{align} 
\nonumber \\
\mathcal{E}_\text{PBC} &= \left\{ dU p(U), |\psi(U)\rangle \right\},  \label{eqn:EPBC} 
\\
\nonumber  \\
\mathcal{E}_\text{OBC}  &= \left\{ dU dU' p(U,U'), |\psi(U,U')\rangle \right\}, \label{eqn:EOBC} \\ \nonumber
\end{empheq}
($dU, dU':$ independent Haar measures on $t$ qubits). In both cases the projected states $|\psi(U)\rangle$ and $|\psi(U,U')\rangle$, which live on the Hilbert space $\mathcal{H}_A$, are  given by $|\psi\rangle = |\Tilde{\psi}\rangle/\Vert |\Tilde{\psi}\rangle \Vert$ with
\begin{align}
\label{eqn:proj_states}
    |\tilde{\psi}(U)\rangle & \equiv \sum_{\sigma \in \{0,1\}^{N_A}} \text{Tr}(W^\sigma U) |\sigma\rangle, \\
    |\tilde{\psi}(U, U')\rangle & \equiv \sum_{\sigma \in \{0,1\}^{N_A}} \langle U| W^\sigma |U'\rangle |\sigma\rangle,
\end{align}
defining the $t$- qubit state $\langle U| \equiv \langle +|^{\otimes t} U^\dagger$ and $|U'\rangle \equiv U'|0\rangle^{\otimes t}$, and they occur with probability densities $p \propto  \| | \tilde{\psi} \rangle \|^2$.
%
Figs.~\ref{fig_2}(b) and (c) give the diagrammatic representations of the states $|\tilde{\psi}(U)\rangle$ and $|\tilde{\psi}(U, U')\rangle$ respectively.


\subsection{Replica trick}

Having derived the projected ensembles that arise in the thermodynamic limit, our next step is to calculate the $k$-th moment operator via Eq.~\eqref{eqn:rho_k}. In the case of PBC, this reads
\begin{align}
\label{eqn:rho_k_PBC}
    \rho^{(k)}_\text{PBC} \propto \int dU \frac{\left(|\tilde{\psi}(U) \rangle \langle \tilde{\psi}(U)|\right)^{\otimes k}}{\langle \tilde \psi(U)| \tilde\psi(U) \rangle^{k-1} },
\end{align}
while in the case of OBC,  
\begin{align}
\label{eqn:rho_k_OBC}
    \rho^{(k)}_\text{OBC} \propto \int\int dU dU' \frac{\left(|\tilde{\psi}(U, U') \rangle \langle \tilde{\psi}(U,U')|\right)^{\otimes k}}{\langle \tilde \psi(U,U')| \tilde\psi(U,U') \rangle^{k-1} }
\end{align}
(the normalization is fixed by enforcing unit trace). 
One sees in both cases that the $k$-th moment operator involves integrals over the unitary group (of $t$ qubits). Now, integrals of polynomial functions of unitaries are typically handled by what is known as Weingarten calculus \cite{collins2006integration,stenberger2021weingarten}; however, in our case, we have a rational function of unitaries, and we hence cannot perform such integrals directly.

To circumvent this obstacle, we employ a replica trick, first introduced by \cite{claeys2022emergent}. We illustrate this for the case of PBC only; the case of OBC follows straightforwardly. Let us instead consider an alternative $k$-moment density operator further indexed by a real number $n$:
\begin{align}
    \rho^{(k,n)}_\text{PBC} & := \int dU p_{k,n}(U) (| \psi(U)\rangle \langle \psi(U) | )^{\otimes k}
\end{align}
where we have modified the probability that each state $|\psi(U)\rangle$ occurs to
\begin{align}
    p_{k,n}(U) = \frac{1}{\mathcal{N}_{k,n}}\langle \tilde{\psi}(U)| \tilde{\psi}(U) \rangle^{k+n},
\end{align}
with the normalization constant $\mathcal{N}_{k,n} =   \int dU \langle \tilde{\psi}(U)| \tilde{\psi}(U) \rangle^{k+n}$.
If we set $n = 1-k$, then $p_{k,1-k}(U) = p(U)$, or in other words
\begin{align}
    \lim_{n \to 1-k} \rho^{(k,n)}_\text{PBC} = \rho^{(k)}_\text{PBC},
\end{align}
i.e., we recover our desired $k$-moment operator.

The virtue of the introduction of the replica operators is that we can write $\rho^{(k+n)}$ in an equivalent form:
\begin{align}
    \rho^{(k,n)}_\text{PBC} = \frac{\int dU \langle \tilde\psi(U)| \tilde\psi(U) \rangle^{n} \left( |\tilde\psi(U)\rangle \langle \tilde\psi(U)| \right)^{\otimes k} }{\int dU \langle \tilde\psi(U)| \tilde\psi(U) \rangle^{k+n}},
    \label{eqn:tilde_rho_nk}
\end{align}
making manifest that the numerator and denominator involve separate {\it independent} integrals over the unitary group (in contrast to Eq.~\eqref{eqn:rho_k_PBC}, which involves a single integral of the ratio over the numerator and denominator). 
At the special values $n \in \mathbb{N}$, $\rho^{(k,n)}$ can thus be tackled using Weingarten calculus as both the numerator and denominator are polynomial functions of $U$.
Our strategy is to thus evaluate $\rho^{(k,n)}_\text{PBC}$ at these integer values, and then formally take the limit $n \to 1-k$ using such information.
We note that strictly speaking, taking a continuous limit from a set of discrete points is a mathematically non-rigorous step\footnote{Another way of putting this is that there are infinitely many ways of extending the domain of a function from the integers to the reals.}.
However, such replica tricks are routinely used in many branches of physics,  such as for calculation of average of logarithm of partition functions in spin glass systems \cite{fischer1993spin} and entanglement entropies in quantum field theories and in context of holographic duality \cite{calabrese2004entanglement,casini2011towards}, among others.
We therefore adopt a similar philosophy in our following analysis, later checking its validity with (unbiased) numerics.

\subsection{Periodic boundary conditions (PBC)}

It suffices to evaluate the numerator $\tilde{\rho}^{(k,n)}_\text{PBC}$ of the expression Eq.~\eqref{eqn:tilde_rho_nk}, a linear operator  which acts on $\mathcal{H}_A^{\otimes k}$, as the denominator of ${\rho}^{(k,n)}_\text{PBC}$ is just the trace of $\tilde{\rho}^{(k,n)}_\text{PBC}$.
For ease of computation, we furthermore consider its vectorization into an element of $(\mathcal{H}_A \otimes \mathcal{H}_A^*)^{\otimes k}$,
\begin{align}
    |\tilde{\rho}^{(k,n)}_\text{PBC} ) := \int dU (\mathbb{I}_{k} \otimes \langle  \mathlarger{\subset}_n | )\left( |\tilde{\psi}(U)\rangle \otimes | \tilde{\psi}(U)\rangle^* \right)^{\otimes (k+n)},
\end{align}
obtained under the Choi-Jamiolkowski map. Above, the identity $\mathbb{I}_{k}$ is the identity map on $(\mathcal{H}_A \otimes \mathcal{H}_A^*)^{\otimes k}$, while $\langle \mathlarger{\subset}_n | = \bigotimes_{i=1}^{n} \left( \sum_{\sigma_i \in \{0,1\}^{N_A}} \langle \sigma_i | \otimes \langle \sigma_i|^*\right)$ is  a $n$-fold tensor product of maximally-entangled states, which lives in the dual space to $(\mathcal{H}_A \otimes \mathcal{H}_A^*)^{\otimes n}$. 
It is best to represent this expression pictorially; Eq.~\eqref{eq:PBC_replica_1} shows the integrand: 
\begin{equation} \label{eq:PBC_replica_1}
     {\includegraphics[width = 0.9\textwidth]{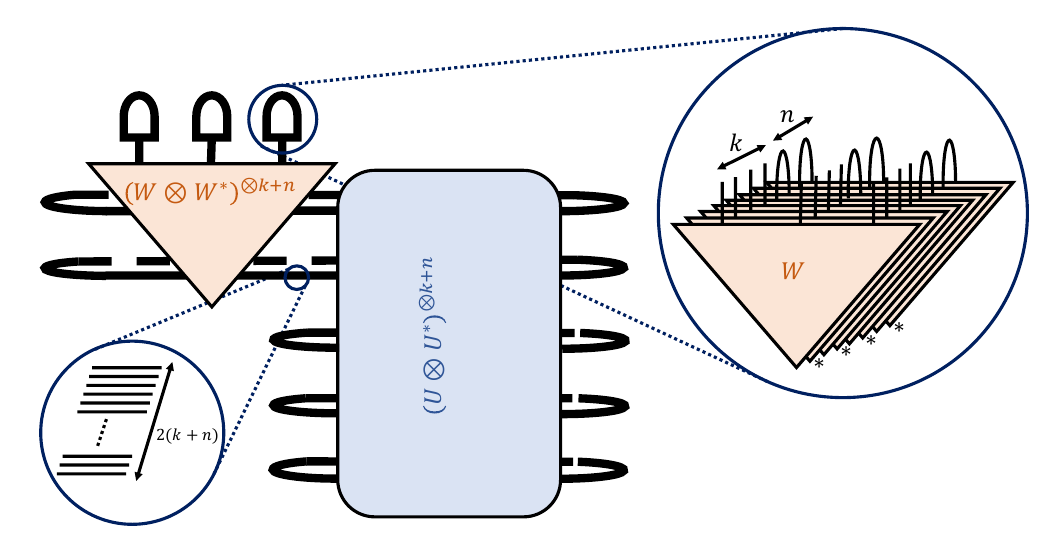}} 
\end{equation} 

We see that the diagram involves $(k+n)$-copies of the unitary pair $U \otimes U^*$ (which will be eventually integrated over the Haar measure), where $U$ acts on $t$ qubits. For each basic unitary $U$ ($U^*$) the inputs and outputs of   $\lceil N_A/2 \rceil$ legs are fed into the tensor $W$ ($W^*$); while the remaining $t - \lceil N_A/2\rceil$ legs are traced over. The tensors $(W \otimes W^*)^{\otimes (k+n)}$ are terminated at the top via a mixture of open legs and `caps', which represent the action of $(\mathbb{I}_{k} \otimes \langle  \mathlarger{\subset}_n | )$, as seen in the zoom-in, and in accordance with our explanation before. 

We are now in a position to perform the integral. The theory of integration over the unitary group tells us that
\begin{align}
\int_{U \sim \text{Haar}(2^t)} dU (U \otimes U^*)^{\otimes (k+n)} = \sum_{\sigma, \tau \in S_{k+n} } W\!g(\sigma \tau^{-1}, 2^t) | P_t(\tau) \rangle \langle P_t({\sigma}) |,
\end{align}
where $S_{k+n}$ is the symmetric group on $k+n$ elements and $W\!g(\sigma,d)$ is the so-called Weingarten function \cite{collins2006integration,stenberger2021weingarten}. 
Here, $|P_t({\sigma})\rangle$ is the vectorization of the permutation operator $P_t(\sigma)$ acting on $k+n$ copies of a $t$-qubit space, i.e., 
\begin{align}
    P_t({\sigma}) & = | i_{\sigma(1)} i_{\sigma(2)} \cdots i_{\sigma(k+n)} \rangle \langle i_1 i_2 \cdots i_{k+n} |, \nonumber \\
    |P_t({\sigma})\rangle & = |i_1 i_{\sigma(1)} i_2 i_{\sigma(2)} \cdots i_{k+n} i_{\sigma(k+n)} \rangle,
\end{align}
where $i \in \{0,1\}^{t}$. In pictures, this leads to the following diagram for $| \tilde{\rho}^{(k,n)}_\text{PBC} )$:
\begin{equation} \label{eq:PBC_replica_2}
     {\includegraphics[width = 0.9\textwidth]{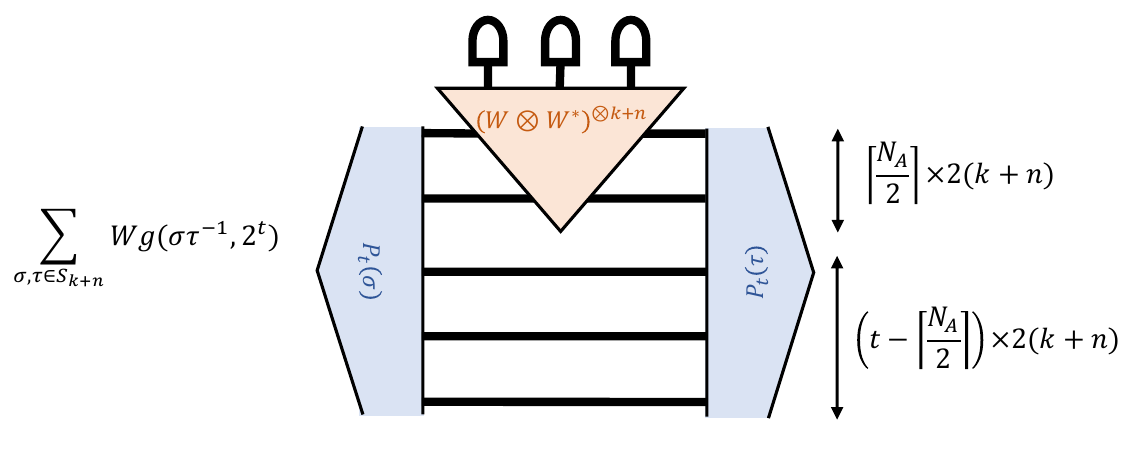}} 
\end{equation}
We can further simplify the figure noting that the contractions involving the part of the diagram with $(t - \lceil N_A/2\rceil)\times (2(k+n))$ qubits (i.e., where $(W \otimes W^*)^{\otimes(k+n)}$ does not directly act) simply involves the inner product 
\begin{align} 
    \langle P_{t-\lceil N_A/2\rceil} (\tau) | P_{t-\lceil N_A/2\rceil} (\sigma) \rangle = \left(2^{t-\lceil N_A/2 \rceil}\right)^{\#(\sigma \tau^{-1})}, 
    \label{eqn:loops}
\end{align}
where $|P_{t-\lceil N_A/2\rceil} (\sigma)\rangle, |P_{t-\lceil N_A/2\rceil} (\tau)\rangle$ are vectorizations of the permutation operator $P_{t-\lceil N_A/2\rceil}(\sigma)$, $P_{t-\lceil N_A/2\rceil}(\tau)$, which each act on $k+n$ copies of $t - \lceil N_A/2\rceil$ qubits, according to permutation elements $\sigma, \tau \in S_{k+n}$. This   evaluates to Eq.~\eqref{eqn:loops}, the dimension raised to $\#(\sigma^{-1} \tau)$, the number of cycles of the permutation element $\sigma \tau^{-1}$. 

This gives us Eq.~\eqref{eq:PBC_perm_ops} as our final simplified diagrammatic expression for $| \tilde{\rho}^{(k,n)}_\text{PBC} )$:
\begin{equation} \label{eq:PBC_perm_ops}
     {\includegraphics[width = 0.7\textwidth]{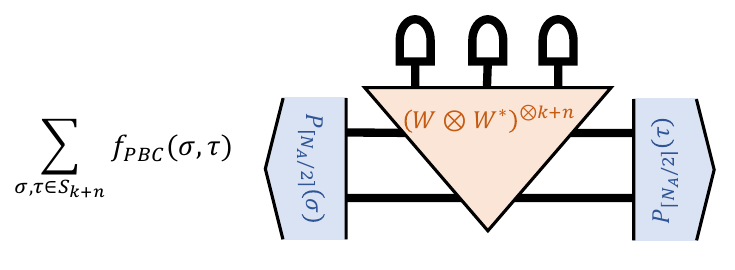}}.
\end{equation}
The important point of the calculation to be appreciated is that the pre-factor in front of the diagrams under the sum,
\begin{align}
    f_{\textrm{PBC}}(\sigma, \tau) \coloneqq  W\!g(\sigma \tau^{-1}, 2^t) \left(2^{t-\lceil N_A/2 \rceil}\right)^{\#(\sigma \tau^{-1})},
\end{align}
 contains {\it all} time-dependency. In   contrast, the  diagrams that remain in Eq.~\eqref{eq:PBC_perm_ops}, while depending on $N_A$, $k$ and $n$, have no   $t$-dependence. 

Referring to Eq.~\eqref{eqn:tilde_rho_nk} and Eq.~\eqref{eq:PBC_perm_ops}, we see that the integral involved in the definition of $\tilde{\rho}^{(k,n)}_\text{PBC}$ has been converted to a double sum over permutation elements $\sigma,\tau$ in $S_{k+n}$.
Our strategy next is to consider separately the diagonal $(\sigma = \tau)$ and off-diagonal $(\sigma \neq \tau)$ contributions.
What we will argue for, is that for any fixed $k$ and fixed $n$ (i) the diagonal contributions are precisely proportional to $|\rho^{(k)}_\text{Haar})$, the vectorization of the $k$-the moment of the Haar ensemble $\rho^{(k)}_\text{Haar}$; and (ii)
the off-diagonal contributions are subleading, in a   parameter that is exponentially small in $t$, compared to diagonal contributions. Specifically, we will argue that any off-diagonal term is at most  $2^{-2t}$ that of the diagonal contribution, asymptotically. Taken together, this will then give us our claim that the limiting distribution at late times is the Haar ensemble, whose convergence in time goes asymptotically as $\sim 2^{-2t}$.

To this end, we first consider the diagonal contributions $\sigma = \tau$. 
In this case the pre-factor is common to all $\sigma$: 
\begin{align}
    f_\text{PBC}(\sigma, \sigma)  = f_\text{PBC}(1,1),
\end{align}
while because $(W \otimes W^*)$ is proportional to an isometry from the temporal slice to the spatial slice  (as mentioned before in Sec.~\ref{sec:analytics}, and detailed in Appendix \ref{sec:W_tensor_contraction})
\begin{align}
    \sum_{i \in \{0,1\}^{\lceil N_A/2\rceil}}   \langle i | \otimes \langle i |^* (W^{\sigma} \otimes W^{*\sigma })  \sum_{j \in \{0,1\}^{\lceil N_A/2\rceil}}   | j \rangle \otimes |j\rangle^* \propto   |\sigma \rangle \otimes |\sigma\rangle^*, 
\end{align}
it can be verified that 
\begin{align}
    \langle P_{\lceil N_A/2\rceil}(\sigma)| W \otimes W^* |P_{\lceil N_A/2\rceil}(\sigma)\rangle = |P_{N_A} (\sigma)\rangle.
\end{align}
 Thus the diagonal contribution simplifies to
\begin{align}
    \sum_{ \sigma \in S_{n+k} } (\mathbb{I}_k \otimes \langle \mathlarger{\subset}_n | ) |P_{N_A} (\sigma)\rangle \propto \sum_{\sigma' \in S_k} |P_{N_A}(\sigma') ) \propto | \rho^{(k)}_\text{Haar} ),
\end{align}
where $P_{N_A}(\sigma) (P_{N_A}(\sigma'))$ is the permutation operator acting on $k+n$ ($k$) copies of $N_A$ qubits according to $\sigma \in S_{k+n}$ ($\sigma' \in S_k$), and $|\rho^{(k)}_\text{Haar})$ is the vectorization of the $k$-th moment of the Haar ensemble $\rho^{(k)}_\text{Haar}$ on $N_A$ qubits.
Importantly, the constants of proportionality in the equation above depend only on $N_A, k, n$ but not $t$.
Thus, we have
\begin{align}
    \text{Diagonal terms}\left( |\tilde{\rho}^{(k,n)}_\text{PBC}) \right) = C_{\text{PBC}}(N_A,k,n) \times  f_\text{PBC}(1,1) \times | \rho^{(k)}_\text{Haar} ).
\end{align}

Next, we move to off-diagonal contributions $\sigma \neq \tau$. In the   limit of large $t$, one can show that the Weingarten function has the following asymptotic behavior \cite{collins2006integration,roberts2017chaos}
\begin{align}
    W\!g(\sigma \tau^{-1},2^t) \sim \frac{1}{(2^{t})^{2(n+k) - \#(\sigma \tau^{-1}) } }
\end{align}
and noting that 
\begin{align}
    \#(\sigma \tau^{-1}) \leq n+k -1
\end{align}
for $\sigma \neq \tau$,  i.e., the number of cycles of a nontrivial permutation element is strictly less than the number of cycles of the identity element, we necessarily have
\begin{align}
    \frac{f_\text{PBC}(\sigma, \tau \neq \sigma)}{f_\text{PBC}(1,1) } \lesssim 2^{-2t},
    \label{eqn:asymptotics1}
\end{align}
that is, in the limit of large $t$, any off-diagonal pre-factor  $f_\text{PBC}$ is at most $2^{-2t}$  that of  the diagonal pre-factor. 

Therefore, we can write $\tilde{\rho}^{(k,n)}$ as
\begin{align}
    \tilde{\rho}^{(k,n)}_\text{PBC} = C_{\text{PBC}}(N_A,k,n) f_\text{PBC}(1,1) \rho^{(k)}_\text{Haar} + \sum_{\sigma \neq \tau \in S_{k+n} } f_\text{PBC}(\sigma,\tau) \tilde{\rho}^{(k,n)}_{\sigma,\tau; \text{PBC}},
\end{align}
where $\tilde{\rho}^{(k,n)}_{\sigma,\tau; \text{PBC}}$ is the $(\sigma,\tau)$ term appearing in the sum of Eq.~\eqref{eq:PBC_perm_ops}, which we note again is independent of $t$.  
We  therefore see that the net  contribution from all off-diagonal terms is subleading to the diagonal contribution,  asymptotically by $2^{-2t}$.

Upon taking the trace of $\tilde{\rho}^{(k,n)}_\text{PBC}$ and normalizing to construct $\rho^{(k,n)}_\text{PBC}$, we can then straightforwardly express the replicated density operator $\rho^{(k,n)}_\text{PBC}$ as 
\begin{align}
    \rho^{(k,n)}_\text{PBC}  = \rho^{(k)}_\text{Haar}  + \delta \rho^{(k,n)}_{\text{PBC}},
\end{align}
where $\delta \rho^{(k,n)}_\text{PBC}$ is traceless and from our analysis, asymptotically $\|\delta \rho^{(k,n)}_\text{PBC}\| \sim 2^{-2t}$ (taken in any norm). Since this is true for any $k,n$, upon taking the limit $n \to 1-k$, the $k$-th moment of the Haar ensemble (at least within the replica trick), has the behavior
\begin{align}\label{PBC_main_result}
    \boxed{ \rho^{(k)}_\text{PBC} = \rho^{(k)}_\text{Haar} + \delta \rho^{(k)}_{\text{PBC}}}
\end{align}
where $\delta \rho^{(k)}_\text{PBC} := \lim_{n \to 1-k} \delta \rho^{(k,n)}_\text{PBC}$, and  asymptotically $\|\delta \rho^{(k)}_{\text{PBC}}\| \sim 2^{-2t}$. This is our first main result.
In particular, this implies that the projected ensemble of the KIM, under periodic boundary conditions, deep thermalizes in the thermodynamic limit exponentially fast in time, at a rate $v_\text{PBC} = -\lim_{t \to \infty} \frac{1}{t} \log_2 \|\delta \rho^{(k)}_{\text{PBC}}\| = 2$.

 \subsection{Open boundary conditions (OBC)}
The numerator from the expression Eq.~\eqref{eqn:tilde_rho_nk}, $\tilde{\rho}^{(k,n)}_\text{OBC}$ in the OBC case becomes (upon vectorization),
\begin{align}
    |\tilde{\rho}^{(k,n)}_\text{OBC} ) := \int dU (\mathbb{I}_{k} \otimes \langle  \mathlarger{\subset}_n | )\left( |\tilde{\psi}(U, U')\rangle \otimes | \tilde{\psi}(U, U')\rangle^* \right)^{\otimes (k+n)}
\end{align}
where the identity $\mathbb{I}_{k}$ and $\langle \mathlarger{\subset}_n | $ are as defined for the PBC case.
For the OBC case, we see that the expression involves $(k+n)$-copies of the state pair $|U\rangle \otimes |U\rangle^*$ and $|U'\rangle \otimes |U'\rangle^*$ (which will be eventually integrated over under the integral) on the two sides of the tensor $(W \otimes W^*)^{\otimes(k+n)}$. For each basic state $|U\rangle$ ($|U\rangle^*$) and $|U'\rangle$ ($|U'\rangle^*$) its   $\lceil N_A/2 \rceil$ legs are fed into the tensor $W$ ($W^*$) on the right and left respectively; while the remaining $t - \lceil N_A/2\rceil$ legs are contracted with each other. Again, the tensors $(W \otimes W^*)^{\otimes (k+n)}$ are terminated at the top via a mixture of open legs and `caps', which represent the action of $(\mathbb{I}_{k} \otimes \langle  \mathlarger{\subset}_n | )$, similar to the PBC case. Pictorially this is shown in Eq.~\eqref{eq:OBC_replica_1},

\begin{equation} \label{eq:OBC_replica_1}
     {\includegraphics[width = 0.7\textwidth]{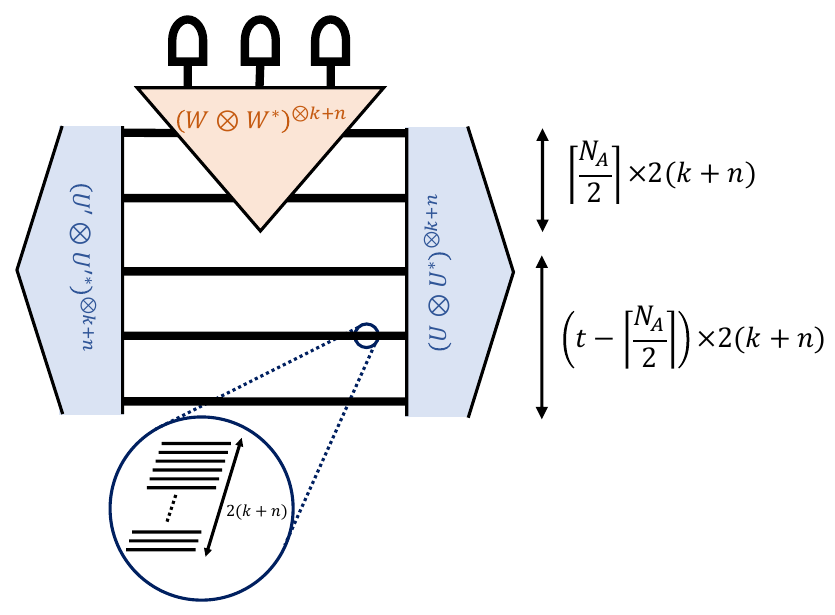}} 
\end{equation}

The appropriate identity from the theory of integration over the unitary group relevant here is now
\begin{align}
\big| \rho_{\text{Haar}(2^t)}^{(k+n)} \big)  &\coloneqq \int dU (|U\rangle\otimes |U\rangle^*)^{\otimes k+n}= \frac{\sum_{\sigma \in S_{k+n}}| P_t(\sigma) \rangle} {2^t(2^t+1)\dots(2^t+k+n-1)}
\end{align}
where $S_{k+n}$ is the symmetric group on $k+n$ elements and $|P_t({\sigma})\rangle$ is the vectorization of the permutation operator $P_t(\sigma)$ acting on $k+n$ copies of a $t$-qubit space as described in the PBC section. As in the PBC case, further simplification follows by noting that the contractions involving the part of the diagram with $k+n$ copies of $t - \lceil N_A/2\rceil$ qubits (i.e., where $(W \otimes W^*)^{\otimes(k+n)}$ does not directly act) simply involves the inner product 
\begin{align}
    \langle P_{t-\lceil N_A/2\rceil} (\tau) | P_{t-\lceil N_A/2\rceil} (\sigma) \rangle = \left(2^{t-\lceil N_A/2 \rceil}\right)^{\#(\sigma \tau^{-1})}, 
\end{align}
where $|P_{t-\lceil N_A/2\rceil} (\sigma)\rangle, |P_{t-\lceil N_A/2\rceil} (\tau)\rangle$ are vectorizations of the permutation operator $P(\sigma), P(\tau)$ which act on $t - \lceil N_A/2\rceil$ qubits, according to permutation elements $\sigma, \tau \in S_{k+n}$.

This finally gives us the figure in Eq.~\eqref{eq:OBC_perm_ops}, 
\begin{equation} \label{eq:OBC_perm_ops}
     {\includegraphics[width = 0.7\textwidth]{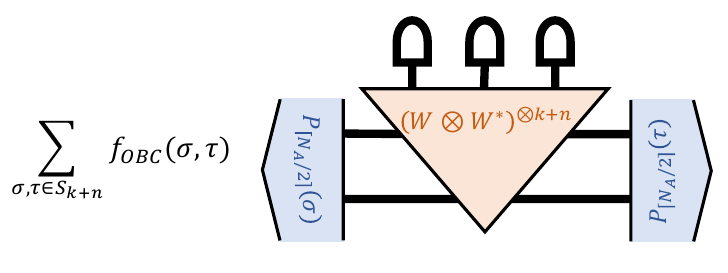}} 
\end{equation}
which we see is identical in form to the corresponding result in the PBC case, Eq.~\eqref{eq:PBC_perm_ops},
except that 
%
the pre-factor in front of the diagrams is  
\begin{align}
    f_{\textrm{OBC}}(\sigma, \tau) \coloneqq  \frac{1}{(2^t(2^t+1)\dots(2^t+k+n-1))^2} \left(2^{t-\lceil N_A/2 \rceil}\right)^{\#(\sigma \tau^{-1})}.
\end{align}
Again, just like in the PBC case, the prefactors $f_\text{OBC}$ contain {\it all} time-dependency; in contrast, the   diagrams in Eq.~\eqref{eq:OBC_perm_ops} have no   $t$-dependence. 

%
%
%

We can thus employ similar logic as in PBC case to consider the relative contributions of the diagonal terms $(\sigma = \tau)$ and off-diagonal-terms $(\sigma \neq \tau)$.
Again for diagonal terms the pre-factor acquires a value that is is common to all $\sigma$: 
\begin{align}
    f_\text{OBC}(\sigma, \sigma)  = f_\text{OBC}(1,1),
\end{align}
and the diagrams 
 simplify to 
\begin{align}
    \sum_{ \sigma \in S_{n+k} } (\mathbb{I}_k \otimes \langle \mathlarger{\subset}_n | ) |P_{N_A} (\sigma)\rangle \propto \sum_{\sigma' \in S_k} |P_{N_A}(\sigma') ) \propto | \rho^{(k)}_\text{Haar} ),
\end{align}
where $P_{N_A}(\sigma) (P_{N_A}(\sigma'))$ is the permutation operator acting on $k+n$ ($k$) copies of $N_A$ qubits according to $\sigma \in S_{k+n}$ ($\sigma' \in S_k$).
Thus, we have
\begin{align}
    \text{Diagonal terms}\left(|\tilde{\rho}^{(k,n)}_\text{OBC})\right) = C_{\text{OBC}}(N_A,k,n) \times  f_\text{OBC}(1,1) \times | \rho^{(k)}_\text{Haar} ).
\end{align}

Next, moving to off-diagonal contributions
$\sigma \neq \tau$, we have, again since $\#(\sigma \tau^{-1}) \leq n+k -1$, the result that asymptotically in large $t$,
\begin{align}
    \frac{f_\text{OBC}(\sigma, \tau \neq \sigma)}{f_\text{OBC}(1,1) } \lesssim 2^{-t}.
    \label{eqn:asymptotics2}
\end{align}
That is, in the limit of large $t$, any off-diagonal pre-factor  $f_\text{OBC}$ is at most $2^{-t}$ as large as  a diagonal pre-factor.
This is to be directly contrasted to the PBC case, Eq.~\eqref{eqn:asymptotics1}, where it was at most twice as small, $2^{-2t}$.

Therefore, we can write $\tilde{\rho}^{(k,n)}$ in the OBC case as
\begin{align}
    \tilde{\rho}^{(k,n)}_\text{OBC} = C_{\text{OBC}}(N_A,k,n) f_\text{OBC}(1,1) \rho^{(k)}_\text{Haar} + \sum_{\sigma \neq \tau \in S_{k+n} } f_\text{OBC}(\sigma,\tau) \tilde{\rho}^{(k,n)}_{\sigma,\tau;\text{OBC}},
\end{align}
where $\tilde{\rho}^{(k,n)}_{\sigma,\tau; \text{OBC} }$ is the $(\sigma,\tau)$ term appearing in the sum of Eq.~\eqref{eq:OBC_perm_ops}, which is independent of $t$.
The replicated density operator $\rho^{(k,n)}$  can thus be expressed 
\begin{align}
    \rho^{(k,n)}_\text{OBC}  = \rho^{(k)}_\text{Haar}  + \delta \rho^{(k,n)}_{\text{OBC}},
\end{align}
where $\delta \rho^{(k,n)}_{\text{OBC}}$ is traceless and asymptotically $\|\delta \rho^{(k,n)}_{\text{OBC}}\| \sim 2^{-t}$. Since this is true for any $k,n$,  upon taking the limit $n \to 1-k$ the $k$-th moment of the Haar ensemble (at least within the replica trick), has the behavior
\begin{align}\label{OBC_main_result}
    \boxed{ \rho^{(k)}_\text{OBC} = \rho^{(k)}_\text{Haar} + \delta \rho^{(k)}_{\text{OBC}}}
\end{align}
where $\delta\rho^{(k)}_\text{OBC} := \lim_{n \to 1-k} \delta \rho^{(k,n)}_\text{OBC}$ and  $\|\delta \rho^{(k)}_{\text{OBC}} \| \sim 2^{-t}$. In particular, this implies that the projected ensemble of the KIM under open boundary conditions, deep thermalizes in the thermodynamic limit also exponentially quickly as in a system with periodic boundary conditions, but with a rate $v_\text{OBC} = -\lim_{t \to \infty} \frac{1}{t} \log_2 \|\delta \rho^{(k)}_{\text{OBC}}\| = 1$ which is seen to be half as small as $v_\text{PBC} = 2$. This is our second main result.

\section{Numerics}
\label{sec:numerics}

\subsection{Replica trick based numerics}

\begin{figure}[t]
 \includegraphics[width=\columnwidth]{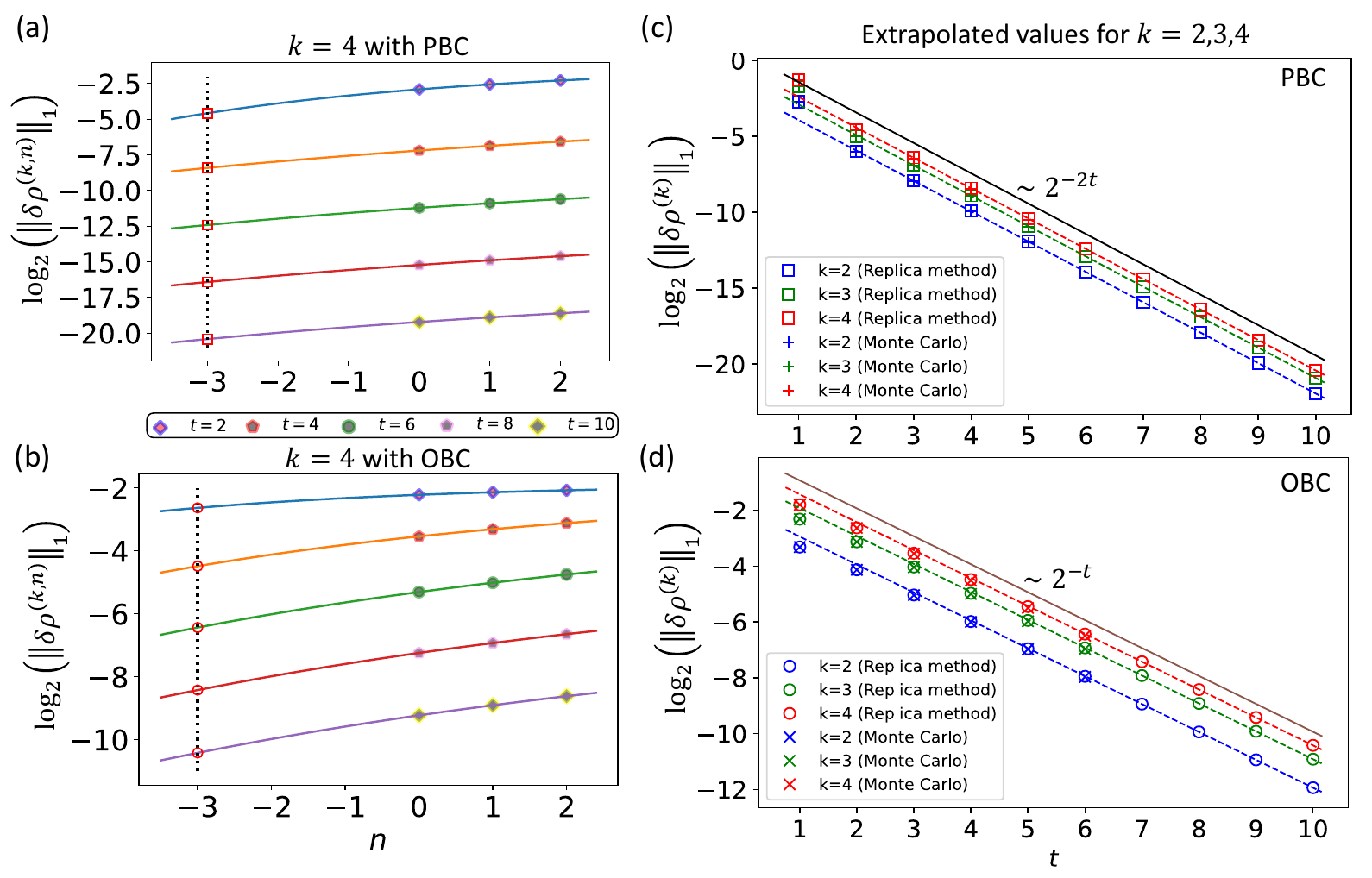}
  \caption{Deviation away from the Haar ensemble, numerically computed within the replica trick and Monte Carlo sampling.
  (a,b) Estimation of deviation $\| \delta\rho^{(k)} \|_1$ from replica trick by evaluating $\delta\rho^{(k,n)} := \rho^{(k,n)} - \rho^{(k)}_\text{Haar}$ using Eqs.~\eqref{eq:PBC_perm_ops} and \eqref{eq:OBC_perm_ops} and extrapolation of $\| \delta \rho^{(k,n)}\|_1$ to $n = 1-k$. Due to computational constraints, we are limited up to $n = 6-k$. Shown in both the (a) PBC case and (b) OBC case are $\| \delta \rho^{(k,n)}\|_1$ for $k = 4$; other moments $k$ behave similarly. In all cases, the data for $\log_2(\| \delta \rho^{(k,n)}\|_1)$ is fitted well by a simple exponential $a_k +b_k e^{-c_k n}$. Extrapolated values are denoted by red hollow squares and circles respectively. 
  (c,d) Deviation $\| \delta\rho^{(k)} \|_1$ from Haar ensemble as a function of time $t$, for the (c) PBC  and (d) OBC cases. 
  Data come from both replica trick numerics as well as Monte Carlo numerics. 
  There is a very good match between these two methods; further, for all moments shown, the deviation behaves  asymptotically as $2^{-2t}$ and $2^{-t}$ respectively (also plotted as solid black and brown lines as guides to the eye), in perfect agreement with our analysis.
  }
  \label{fig:nums_replica}
\end{figure}

We support our main analytical results regarding the asymptotic approach to deep thermalization, Eqs.~\eqref{PBC_main_result} and \eqref{OBC_main_result}, with numerics based on the replica operator $\rho^{(k,n)}$.
Specifically, we numerically construct, for both PBC and OBC, the unnormalized moment operator $\tilde{\rho}^{(k,n)}$ 
following Eqs.~\eqref{eq:PBC_perm_ops} and \eqref{eq:OBC_perm_ops}, for subsystem size $N_A = 2$, moments $k = 2,3,4$, different times $t$, and various $n$.
$\rho^{(k,n)}$ is then obtained upon normalizing the resulting operator to have unit trace. We then define $\delta \rho^{(k,n)} \coloneqq \rho^{(k,n)} - \rho_{\text{Haar}}^{(k)}$ (for both PBC and OBC) as the deviation away from the Haar ensemble, and plot, for a fixed $k$ and $t$ its trace norm $\|\delta \rho^{(k,n)} \|_1$ for various values of $n$.
Results for $k = 4$ and various $t$ are shown in Fig.~\ref{fig:nums_replica}(a,b); similar trends are observed for other moments $k$ (not shown).
We find that $\log_2 \|\delta \rho^{(k,n)} \|_1$ is always well fitted by a simple exponential function   $a_k + b_k e^{-c_k n}$ for some real numbers $a_k,b_k,c_k$; from this fit we extrapolate to find $\|\delta \rho^{(k,1-k)} \|_1$ which defines the deviations $\|\delta \rho^{(k)}_{\text{PBC}}\|_1$ and $\|\delta \rho^{(k)}_{\text{OBC}}\|_1$ at a given $k$ and $t$ (red squares and circles in Fig.~\ref{fig:nums_replica}(a,b) respectively).
Finally, for a given moment $k$ we plot these extrapolated values as a function of $t$, as shown in Fig.~\ref{fig:nums_replica}(c). 
One sees from the figure that asymptotically  $\| \delta \rho^{(k)}_\text{PBC} \|_1$ and $\| \delta \rho^{(k)}_\text{OBC} \|_1$ follow very well the trend $2^{-2t}$ and $2^{-t}$ respectively, confirming our analysis of their asymptotic behavior in time.

\subsection{Numerics based on Monte Carlo sampling}
The numerics performed in the previous subsection involved simulations confirming our analysis within the replica trick. However, there is an ambiguity in the replica trick, as the fit used to find $\|\delta \rho^{(k)}\|_1 := \|\delta \rho^{(k,1-k)}\|_1$ from the data $\|\delta \rho^{(k,n)}\|_1$ was not unique. 
To check the replica method, we construct the moment operators of the projected ensembles from first principles, i.e., from Eqs.~\eqref{eqn:rho_k_PBC} and \eqref{eqn:rho_k_OBC} and  the definition of the projected states Eq.~\eqref{eqn:proj_states}, sampling the unitaries $U, U'$ from the Haar measure over $t$ qubits independently.  
Precisely, we form the unbiased estimators
\begin{align}
    \rho'^{(k)}_\text{PBC} = \frac{1}{M} \sum_{i=1}^M \mathcal{C}_\text{PBC} \frac{\left(|\tilde{\psi}(U_i) \rangle \langle \tilde{\psi}(U_i)|\right)^{\otimes k}}{\langle \tilde \psi(U_i)| \tilde\psi(U_i) \rangle^{k-1} }
\end{align}
and 
\begin{align}
    \rho'^{(k)}_\text{OBC} = \frac{1}{M} \sum_{i=1}^M \mathcal{C}_\text{OBC} 
 \frac{\left(|\tilde{\psi}(U_i, U_i') \rangle \langle \tilde{\psi}(U_i,U_i')|\right)^{\otimes k}}{\langle \tilde \psi(U_i,U_i')| \tilde\psi(U_i,U_i') \rangle^{k-1} },
\end{align}
where $i$ indexes a sample run, $M$ represents the total number of samples simulated, and $\mathcal{C}_\text{PBC/OBC}$ is a constant of proportionality ensuring $\rho^{(k)}_\text{PBC/OBC}$ in Eqs.~\eqref{eqn:rho_k_PBC} and \eqref{eqn:rho_k_OBC} has unit trace. In the limit of large $M$, $\rho'^{(k)}_\text{PBC/OBC}$ will converge to $\rho^{(k)}_\text{PBC/OBC}$; we use the converged value of $\| \rho'^{(k)}_\text{PBC/OBC} - \rho^{(k)}_\text{Haar} \|_1$ as an approximation for the true deviation $\| \delta \rho^{(k)} \|_1$ from the Haar ensemble (see Appendix \ref{sec:MC_numerics} for details on the numerics).
For various $t$ we   plot the results on Fig.~\ref{fig:nums_replica}(c). We see that the data from the replica numerics matches very well with the data from the unbiased Monte Carlo numerics, validating the replica method.

\clearpage
\section{Discussion}
\label{sec:discussion}
In this work, we have investigated the role of global system topology in governing the rate of deep thermalization in chaotic quantum many-body systems. 
Specifically, for dynamics under the maximally-chaotic (1+1)d kicked Ising model, we find that while regular thermalization of a small contiguous region    located deep in the bulk  to infinite temperature   requires  time $t_1 = \lceil N_A/2\rceil$, regardless of boundary conditions, the situation for deep thermalization is rather different:
we have derived that the deviation $\Delta^{(k)}$ of the projected ensemble from the maximally-entropic uniform wavefunction distribution behaves asymptotically in time as $2^{-2t}$ and $2^{-t}$, for a system with periodic and open boundary conditions respectively (up to multiplicative factors that may depend on the moment). 
This implies that the deep thermalization times $t_{k \geq 2}$ obey
\begin{align}
     \frac{ t_{k,\text{PBC}} }{t_{k,\text{OBC}}} =  \frac{1}{2} \qquad (k \geq 2),
    \label{eqn:tk_separation}
\end{align}
as the precision $\varepsilon \to 0$ (recall the meaning of $\varepsilon$ in Eq.~(\ref{eqn:deep_thermalization_times})). 
The distinct equilibration times depending on boundary conditions highlights deep thermalization's non-local nature, in contrast to regular thermalization, which only depends on the build-up of entanglement between a local region and its surroundings, a physical process constrained by information propagation bounds.



From a technical standpoint, we note that our results, Eqs.~\eqref{PBC_main_result}, \eqref{OBC_main_result} and \eqref{eqn:tk_separation}, extend
straightforwardly to dynamics under generic dual-unitary circuits --- that is, circuits comprised of two-local elementary gates which are unitary under a space-time rotation. 
This is because all steps involved in our derivation  depend only on the KIM's dual-unitary nature, and is thus actually not specific to the KIM itself (see Ref.~\cite{ippoliti2022solvable} for the precise conditions for which dual-unitary circuits our results generalize to). 

More broadly, one might wonder to what extent a separation of deep thermalization timescales depending on boundary conditions  persists in systems with more general dynamics beyond dual-unitary circuits. 
For quantum circuit models which are still in (1+1)d, it has been argued in Ref.~\cite{ippoliti2023dynamical} that dynamical purification --- that is, the phenomenon of a mixed state eventually purifying under monitored quantum dynamics (i.e., dynamics with entangling gates and measurements), when applied to the space-time dual of the circuit, leads to an effective constrained size of the projected ensemble at finite time, even in the thermodynamic limit. That is, even though the bath being measured is infinitely large leading in principle to an infinite number of projected states, there are only so many non-degenerate states that contribute:  specifically, those states labeled by measurement outcomes within some large but finite correlation length around the subregion $A$ of interest (the measurement information outside is lost due to dynamical purification). In our context, this would translate to the fact that at finite time, there is only a finite neighborhood surrounding region $A$ which matter  for the purposes of the projected ensemble, thereby potentially washing out the effects of boundary conditions. This is much like the ballistic light-cone outside of which correlations and quantum entanglement cannot spread beyond, which constrains the regular thermalization time and renders  effects of boundary conditions on dynamics at any finite time irrelevant. However, we stress the physical origin of the two is distinctly different: the former arises from purification dynamics, while the latter comes from the Lieb-Robinson velocity of information propagation. 
In systems of higher dimensions though, the physics of dynamical purification is not expected to be relevant (see \cite{ippoliti2023dynamical} and \cite{Bao}), such that it may be expected that once again deep thermalization rates may  depend distinctly on the  boundary conditions of the system. We leave  the precise quantitative explorations of such interesting physics, to future work.

\section*{Acknowledgements}


We thank Matteo Ippoliti, Tibor Rakovszky, Tianci Zhou, and Wai-Keong Mok for useful discussions. 
We acknowledge support from the National Research Foundation (NRF) Singapore  and the Ministry of
Education (MOE), Singapore.
HS acknowledges funding from the Centre for Quantum Technologies (CQT) PhD program. 
WWH is supported by the NRF Fellowship, NRF-NRFF15-2023-0008.
Numerical simulations were performed using the National University of Singapore (NUS)'s high-performance computing facilities, partially supported by NUS IT’s Research Computing group.

\begin{appendix}
\section{Construction of $W$ tensor from $W'$ tensor}
\label{sec:W_tensor}

We describe here how to obtain the $W$ tensor from the $W'$ tensor, which appear in the construction of the unnormalized projected state for  finite size systems  and for thermodynamically large systems respectively. 
We remind the reader these tensors are diagrammatically defined in Fig.~\ref{fig_2}(a) and Fig.~\ref{fig_2}(b,c) of the main text. 

As mentioned in the main text, $U(z_1)$ and $U(z_2)$ of Fig.~\ref{fig_2}(a), when considered over all $z_1, z_2$, can in the thermodynamic limit be replaced by independent Haar random unitaries on $t$ qubits.
Now the Haar measure has the following property of being left and right invariant: 

\begin{equation}
    \int_{U \sim \textrm{Haar}} dU f(U) = \int_{U \sim \textrm{Haar}} dU f(UV) = \int_{U \sim \textrm{Haar}} dU f(VU),
\end{equation}
and further the tensor $W'$   can be decomposed as
\begin{equation}
    W'^{\sigma} = U_2 W^{\sigma} U_1, \quad \text{ with } \sigma \in \{0,1\}^{\times N_A},
\end{equation}
allowing us to `absorb' the unitaries $U_1, U_2$ under the integral. 
Thus, the unnormalized projected states are given by Eq.~\eqref{unnorm_1} as stated in the main text. 

\begin{figure}[t]
  \includegraphics[width=\columnwidth]{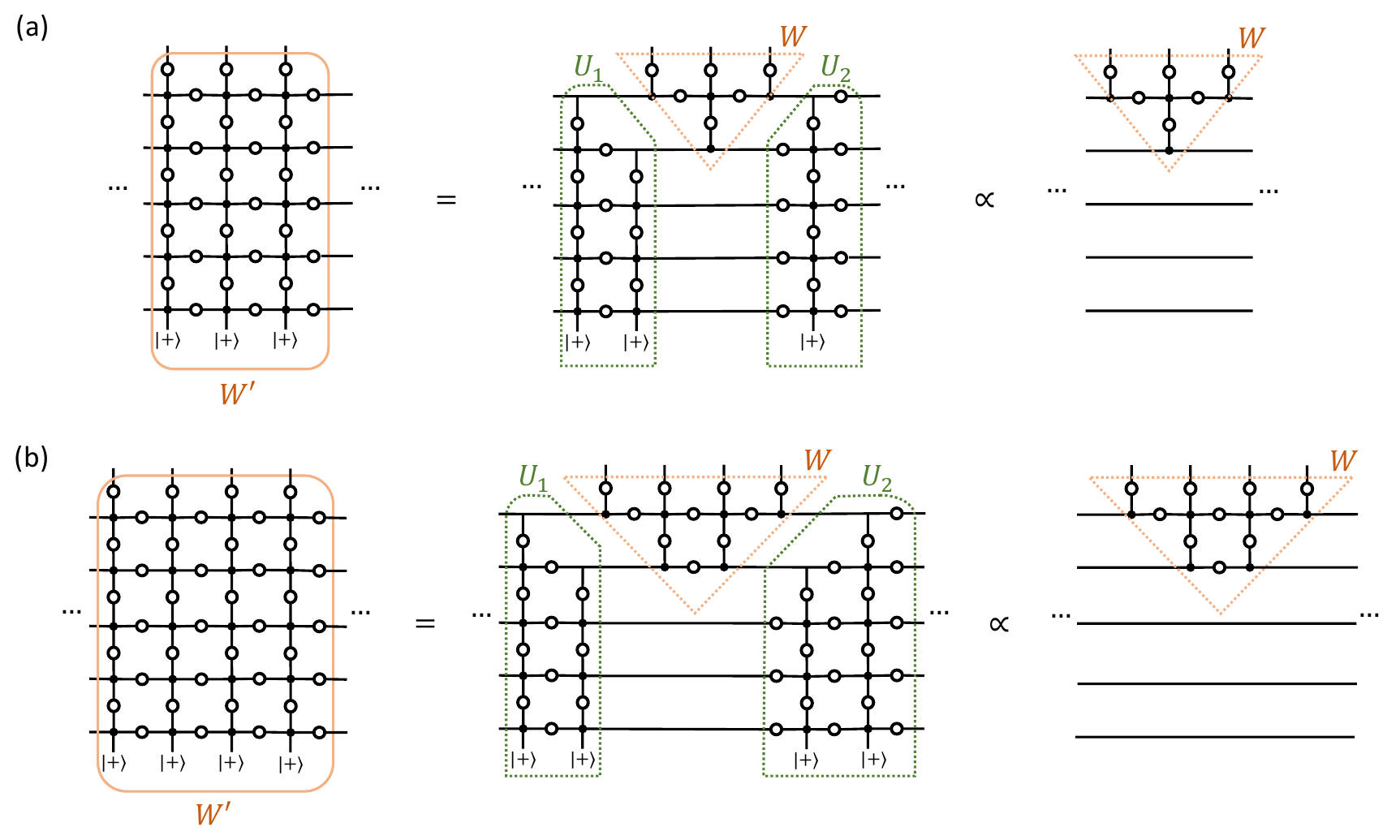}
  \caption{Decomposition of tensor $W'$ into unitaries $U_1, U_2$ (green boxes, up to proportionality factors) as well as tensor $W$. This is shown for representative (a) odd $N_A$ and (b) even $N_A$ cases. In both cases, $W$ involves a `triangle' of elementary tensors, but the precise layout of these tensors differ slightly depending on the oddness/evenness of the subsystem size.  The unitaries $U_1, U_2$ can be absorbed by the Haar random unitaries on the left and right of the figure (not shown). 
  }
  \label{W'_W_contraction}
\end{figure}

The decomposition of $W'$ as mentioned above is shown pictorially in Fig.~\ref{W'_W_contraction} for two representative subsystem sizes $N_A$, showcasing the slight difference in construction of $W$ for the odd and even cases.


\section{Contraction of a $(W,W^*)$ pair}
\label{sec:W_tensor_contraction}
\begin{figure}[t]
\centering
  \includegraphics[width=0.75\columnwidth]{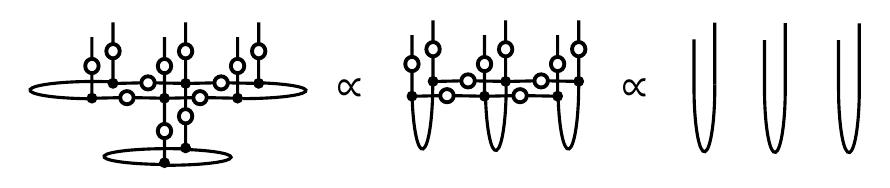}
  \caption{Contraction of the tensor $W$ contacted with $W^{*}$ along the spatial direction yields a term proportional to the identity tensor over $N_A$ qubits. 
  This is denoted by the `cups' (products of Bell pairs) in the right-most diagram, here illustrated for $N_A=3$.}
  \label{WW_pair_contraction}
\end{figure}

Using the rules of diagrammatic manipulations laid out in Sec.~\ref{sec:analytics}, the contraction of a single $W$ tensor with a single $W^*$ tensor along the spatial direction can be easily seen to be proportional to the  identity matrix on the local subsystem $A$, as shown in Fig.~\ref{WW_pair_contraction}. %
This demonstrates the isometric condition $\sum_{\tau \tau'} W^\sigma_{\tau \tau'} W^{* \sigma}_{\tau \tau'} \propto \delta_{\sigma \sigma'}$, as claimed in the main text. Though we show this only for the odd $N_A$ case, it can be easily seen that the diagrams for the even $N_A$ case behave similarly. 


\section{Additional details on Monte Carlo numerics}
\label{sec:MC_numerics}
\begin{figure}[t]
 \includegraphics[width=\columnwidth]{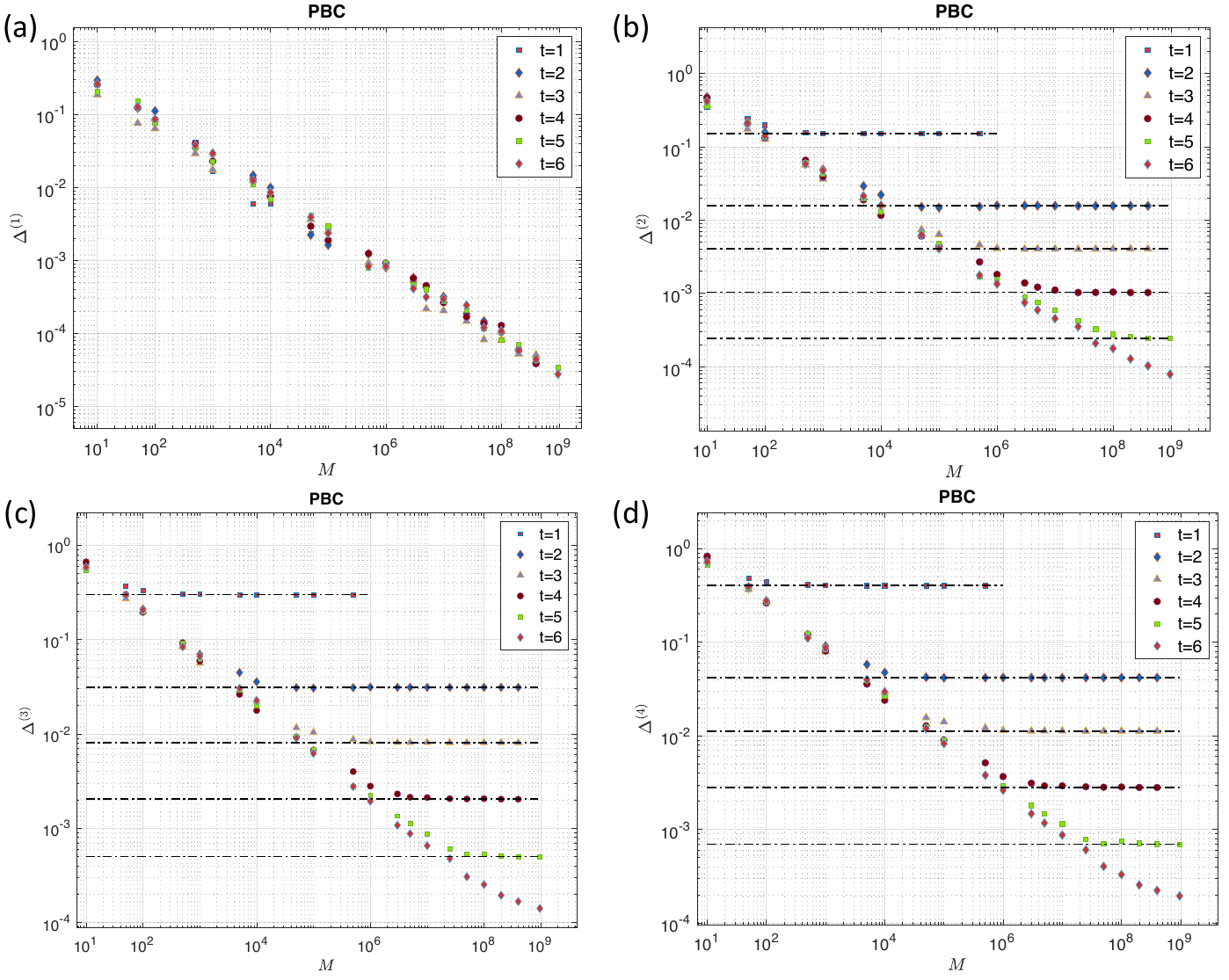}
  \caption{Trace distance  $\Delta^{(k)}_{\text{PBC}}$
  versus total number of samples $M$. 
  The converged values at a given $t$ for $k \geq 2$ are obtained by the average of the last three points of any data series.
  %
  Note that at $M = 10^9$ we have not obtained satisfactory convergence of $\Delta^{(k)}_{\text{PBC}}$ for time $t=6$.}
  \label{fig:PBC_HPC_nums}
\end{figure}

\begin{figure}[t]
 \includegraphics[width=\columnwidth]{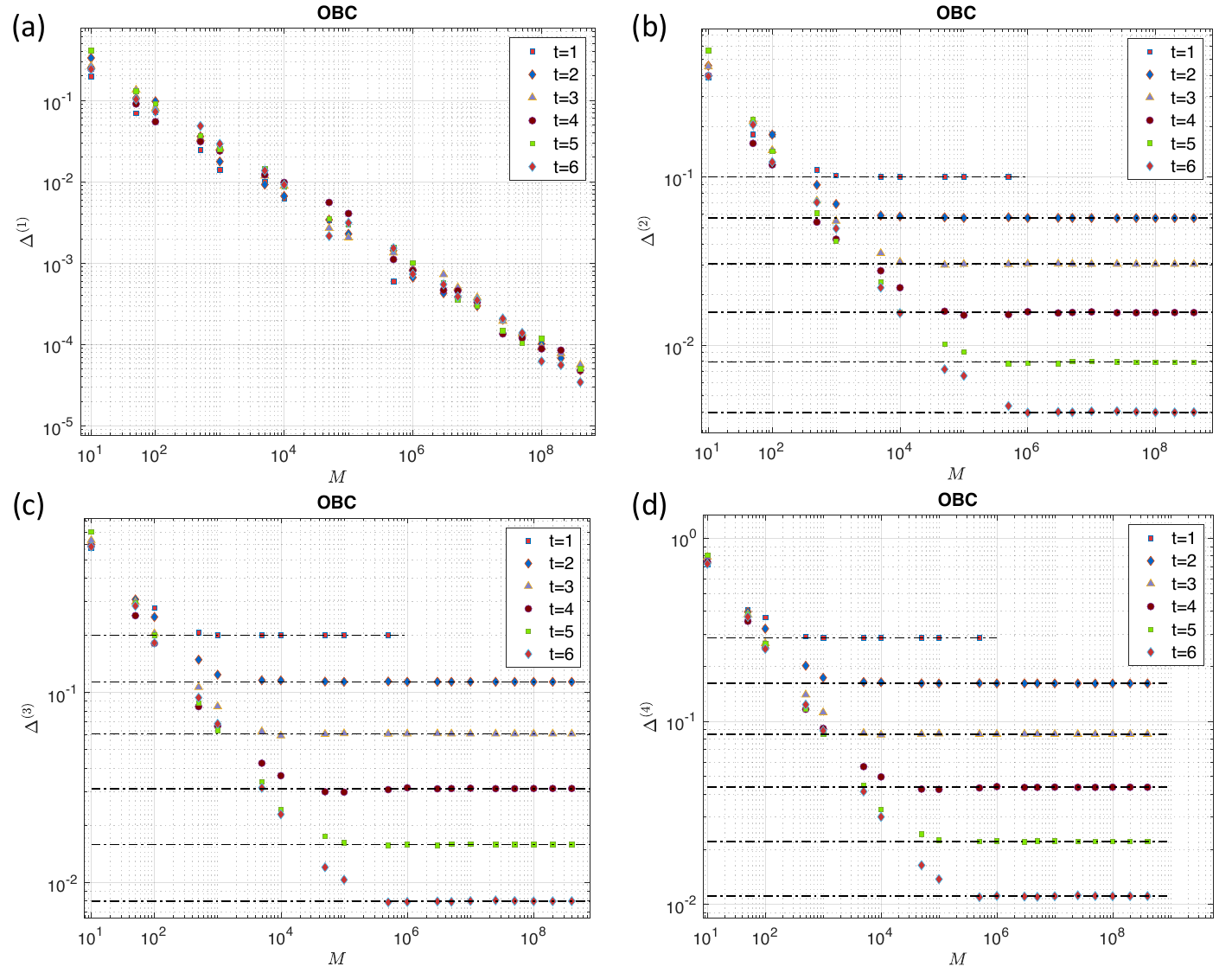}
  \caption{Trace distance  $\Delta^{(k)}_{\text{OBC}}$
  versus total number of samples $M$. 
  The converged values at a given $t$ for $k \geq 2$ are obtained by the average of the last three points of any data series.}
  \label{fig:OBC_HPC_nums}
\end{figure}

In Figs.~\ref{fig:PBC_HPC_nums} and \ref{fig:OBC_HPC_nums} we plot the trace distance of the unbiased estimator to the Haar ensemble $\Delta^{(k)} := \| \rho'^{(k)} - \rho^{(k)}_\text{Haar} \|$ versus total sample number $M$, for the subsystem size $N_A = 2$. 
We see that for $k = 1$ and for $t \geq \lceil N_A/2 \rceil$, the distance decreases monotonically without bound, in agreement with the fact the reduced density matrix is (provably) maximally mixed. 
For $k \geq 2$, the trace distance similarly decreases, but at large enough $M$, it is seen to converge. We extract the converged values through the average of the last three points of the data series. 

\end{appendix}




\clearpage
\bibliography{main.bib}

\nolinenumbers

\end{document}